\documentclass[twocolumn]{aastex62}

\usepackage[caption=false]{subfig}
\usepackage{natbib}
\usepackage{graphicx}
\usepackage{times}
\usepackage{amsmath,amstext}

\newcommand{\ugri}{$ugri$ }
\newcommand{\griz}{$griz$ }

\LTcapwidth=\textwidth
\begin{document}
%\accepted{for publication in ApJ}
%\submitjournal{AASJournal name}


\title{The Space Density of Intermediate Redshift, Extremely Compact, Massive Starburst Galaxies}

\author{Kelly E. Whalen}
\affil{Department of Physics and Astronomy, Dartmouth College, Hanover, NH 03755, USA}

\author{Ryan C. Hickox}
\affil{Department of Physics and Astronomy, Dartmouth College, Hanover, NH 03755, USA}

\author{Alison L. Coil}
\affil{Center for Astrophysics and Space Sciences, University of California, San Diego, La Jolla, CA 92093, USA}

\author{Aleksandar M. Diamond-Stanic}
\affil{Department of Physics and Astronomy, Bates College, Lewiston, ME, 04240, USA}

\author{James E. Geach}
\affil{Centre for Astrophysics Research, University of Hertfordshire, Hatfield, Hertfordshire AL10 9AB, UK}

\author{John Moustakas}
\affil{Department of Physics and Astronomy, Siena College, Loudonville, NY 12211, USA}

\author{Gregory H. Rudnick}
\affil{Department of Physics and Astronomy, University of Kansas, Lawrence, KS 66045, USA}

\author{David S.~N. Rupke}
\affil{Department of Physics, Rhodes College, Memphis, TN, 38112, USA}

\author{Paul H. Sell}
\affil{Department of Astronomy, University of Florida, Gainesville, FL, 32611 USA}

\author{Christy A. Tremonti}
\affil{Department of Astronomy, University of Wisconsin-Madison, Madison, WI 53706, USA}

\author{Julie D. Davis}
\affil{Department of Astronomy, University of Wisconsin-Madison, Madison, WI 53706, USA}

\author{Serena Perrotta}
\affil{Center for Astrophysics and Space Sciences, University of California, San Diego, La Jolla, CA 92093, USA}

\author{Grayson C. Petter}
\affil{Department of Physics and Astronomy, Dartmouth College, Hanover, NH 03755, USA}

\accepted{for publication in ApJ}

 \shortauthors{Whalen et al.}
 \shorttitle{$z\sim0.5$ Compact Starburst Space Density}

\begin{abstract}
We present a measurement of the intrinsic space density of intermediate redshift ($z\sim0.5$), massive ($M_{*} \sim 10^{11} \ \text{M}_{\odot}$), compact ($R_{e} \sim 100$ pc)  starburst ($\Sigma _{SFR}  \sim 1000 \ \text{M}_{\odot} \ \text{yr}^{-1} \text{kpc}^{-1}$) galaxies with tidal features indicative of them having undergone recent major mergers. A subset of them host kiloparsec scale, $>1000 \ \text{km}\ \text{s}^{-1}$ outflows and have little indication of AGN activity, suggesting that extreme star formation can be a primary driver of large-scale feedback. The aim for this paper is to calculate their space density so we can place them in a better cosmological context. We do this by empirically modeling the stellar populations of massive, compact starburst galaxies. We determine the average timescale for which galaxies that have recently undergone an extreme nuclear starburst would be targeted and included in our spectroscopically selected sample. We find that massive, compact starburst galaxies targeted by our criteria would be selectable for $\sim 148 ^{+27}_{-24}$ Myr and have an intrinsic space density $n_{\text{CS}} \sim (1.1^{+0.5}_{-0.3}) \times 10^{-6} \ \ \text{Mpc}^{-3}$. This space density is broadly consistent with our $z\sim0.5$ compact starbursts being the most extremely compact and star forming low redshift analogs of the compact star forming galaxies in the early Universe as well as them being the progenitors to a fraction of intermediate redshift post starburst and compact quiescent galaxies.

\end{abstract}


\section{Introduction}
Galaxy formation models within a $\Lambda$-Cold Dark Matter ($\Lambda$CDM) framework that do not include feedback typically over-predict the present day baryon fraction as well as the number of number density of galaxies on the high and low mass ends of the local stellar mass function (SMF) \citep[e.g.,][]{crot06sagalev,kere09LCDMfeedback,most10hm-sm,mous13primus}. This implies that star formation over cosmic timescales is inefficient, which requires that galaxy formation models inject energy into cooling clouds of gas. This is typically done by invoking feedback from massive stars and active galactic nuclei (AGNs) to heat and eject gas, thus reducing star formation efficiency \citep[e.g.,][]{sprin05galform,dima05quasarfeedback,some15formationrev}. Feedback as a driver of the cosmic star formation inefficiency is supported by evidence of large-scale gas outflows and/or relativistic jets in star forming and active galaxies \citep[e.g.][]{veil05galwindsrev,mcna07AGNfeedback,fabi12AGNfeedback,some15formationrev}.

In massive galaxies, feedback-driven outflows are often attributed to AGN activity since dark matter halo mass, galaxy stellar mass, bulge mass, and black hole mass all scale with one another \citep[e.g.,][]{ferr00scalingRelations,guo10mass,korm13coevolution}. However, cosmological galaxy formation simulations show that the exclusion of stellar feedback in models leads to the formation of galaxies that are $\sim 10$ times more massive than observed at a given redshift, showing that stellar-driven feedback plays an integral role in regulating star formation in massive galaxies \citep[e.g.,][]{sprin05galform, hopk12stellarfeedback}. On small (giant molecular cloud) scales, feedback can slow the local star formation rate by decreasing the gas surface density in a region, but this alone is not sufficient to produce simulated galaxies whose masses match those observed. Large-scale galactic wind-driven outflows where $\dot{M}_{*, outflow} \sim \text{SFR}$ are necessary to be able to model galaxies with masses that are consistent with observations \citep[e.g.,][]{veil05galwindsrev}. 

Constraining the importance of feedback-driven quenching is crucial to understanding how massive galaxies form, especially at high redshift. Massive, quiescent galaxies at $z>1.5$ are typically more compact than their local counterparts by roughly a factor of 5 \citep[e.g.][]{zirm05highzcompact, vand08highzcompact,vand14highzcompact}. The likely progenitors of these massive, compact quiescent galaxies are similarly compact star forming galaxies that were formed in gas-rich mergers of disk galaxies and were then rapidly quenched via some dissipative feedback \citep[e.g.,][]{barr13CANDELScompact,stef13compactprogen, vand15massivecompact}. However, heavy dust obscuration coupled with high redshift makes constraining the role of AGN vs. stellar-driven feedback difficult with the typical UV signatures of outflows \citep[e.g.,][]{vand15massivecompact}.

\begin{figure}
\centering
\includegraphics[width=0.5\textwidth]{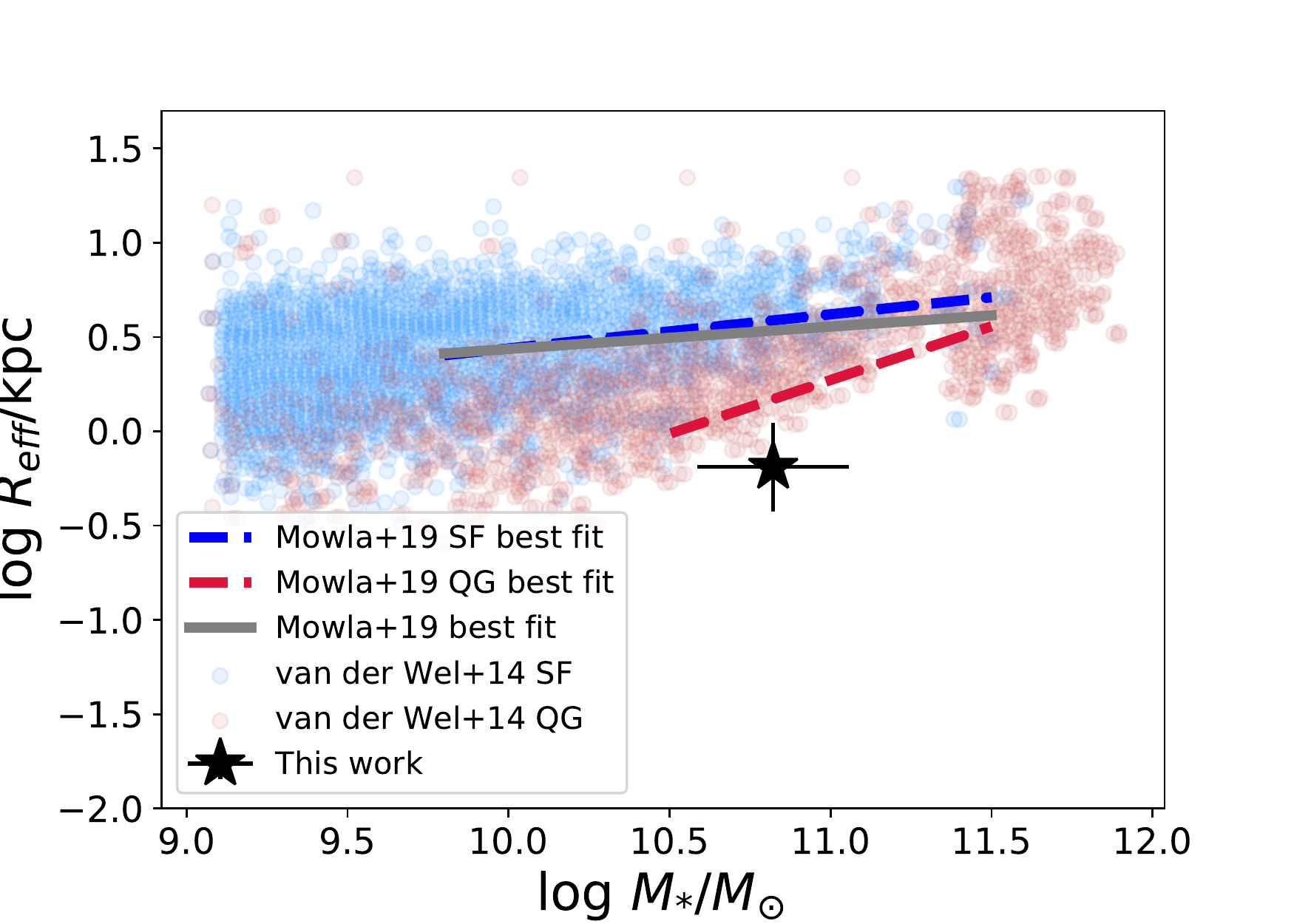}
\caption{Location of our galaxies (black star) within the $0.5 < z < 1$ size-mass plane as presented in \citet{mowl19sizemass}. Blue and red points are \citet{vand14highzcompact} star forming  and quiescent galaxies, respectively. The red, blue, and grey lines are the best fit size-mass relations for the quiescent, star forming, and total CANDELS/3DHST galaxies in \citet{mowl19sizemass}. Our data point represents the average $R_{eff}$ and $M_{*}$ for a subset of the MgII galaxies presented in \citet{davi_inprep_mgii_winds}. Our sources are significanly more compact than other galaxies at similar $z$ and $M_{*}$.}
\label{fig:size_mass}
\end{figure}

We have been studying a population of $z\sim 0.5$ massive, compact galaxies which show signs of recent, extreme bursts of star formation and gas depletion, similar to what we would expect as the progenitors to high-$z$ massive, quiescent galaxies \citep{trem07CSoutflow,diam12CSstarform,diam21CSoutflows,geac13CSmolec,sell14CSoutflow,geac14CSstellarfeedback,rupk19makani,pett20radio}. Our sample of galaxies consists of sources initially targeted as SDSS quasars, but subsequently classified as young post-starburst galaxies due to their blue stellar continua, weak nebular emission lines, and bright infrared photometry \citep{trem07CSoutflow}. Hubble Space Telescope (\textit{HST}) imaging showed that these galaxies have extremely compact morphologies ($R_{e} \sim 100$ pc) with tidal features indicative of having recently undergone a major merger event (see Figure \ref{fig:cutouts}) \citep{diam12CSstarform,sell14CSoutflow}.  We also note that rings and diffraction spikes from the \textit{HST} PSF are visible in the images of our sources, showing that their angular sizes are on the order of that of the PSF which further highlights their compactness \citep{sell14CSoutflow,diam21CSoutflows,davi_inprep_mgii_winds}. The sources in our sample can have SFR surface densities up to $\sim 1000 \ \text{M}_{\odot} \ \text{yr}^{-1} \text{kpc}^{-1}$ \citep{diam12CSstarform,sell14CSoutflow},{ and lie below the $0.5<z<1$ size-mass relations for star forming and quiescent galaxies (see Figure \ref{fig:size_mass}; \citealt{mowl19sizemass,diam21CSoutflows})}. Spectroscopic observations show that these galaxies host outflows with velocities $> 1000 \ \text{km} \ \text{s}^{-1}$ that can extend to tens of kpc \citep{trem07CSoutflow,rupk19makani,davi_inprep_mgii_winds}. There is also little evidence that these massive outflows are primarily driven by AGN activity based on X-ray, IR, radio, and spectral line diagnostics, meaning that extreme star formation can be responsible for gas depletion in these galaxies \citep{diam12CSstarform,sell14CSoutflow,pett20radio}.

These galaxies are important because they allow us to directly observe the effects of extreme star formation on gas kinematics in starburst and post-merger galaxies. In merger-driven galaxy evolution scenarios, a major merger event can trigger a strong burst of obscured star formation. Dissipative feedback via AGN or starburst activity can then expel large amounts of gas and dust from the galaxy, allowing it to passively evolve into a gas-poor massive elliptical galaxy \citep[e.g.][]{sand88ULIRGs,lons06ulirg}. The objects we are studying can possibly be representative of galaxies that are actively undergoing quenching, and might be an important phase for the building up of a massive, quiescent elliptical population. However, this is difficult to determine without knowing the space density of extreme compact starburst galaxies like the ones we have been studying. We are broadly defining our compact starbursts as massive, centrally concentrated galaxies that have recently experienced a burst of star formation. The space density of extreme massive, compact starbursts is strongly dependent on the timescales upon which starburst events can be observed using our selection criteria. 

The aim of this paper is to estimate the average amount of time sources in a simulated galaxy population would be selected as extreme compact starburst galaxies under our selection criteria, in addition to their space density. We also place our galaxies into context with their high redshift compact star forming analogs, compact quiescent galaxies, post starburst galaxies, ultraluminous infrared galaxies (ULIRGs), the merger rate density, and massive, quiescent galaxies within the same redshift interval \citep[e.g.][]{sand88ULIRGs,lons06ulirg,lotz11mergerrate,barr13CANDELScompact,vand14highzcompact,wild16PSBs}.

The outline of the paper is as follows: in Section 2 we discuss the selection of the parent sample of galaxies. In Section 3 we discuss empirical model construction and constraining model free parameters via an MCMC routine. In Section 4 we discuss our implementation of the SDSS quasar selection function. In Section 5 we calculate the average observability timescale and space density for our population of compact starbursts. In Section 6 we place our galaxies into cosmological context with other phases of merger-driven galaxy evolution. We adopt a cosmology of $H_{0} = 70.2 \ \textrm{km} \textrm{s}^{-1} \textrm{Mpc}^{-1}$,
$\Omega _{M} = \Omega _{CDM} + \Omega _{b} = 0.229+0.046 = 0.275$, and
$\Omega _{\Lambda} = 0.725$ \citep{koma11WMAP}

\section{The observed sample}
The selection criteria used for our sample will be detailed in Tremonti et al. in prep, but we will give a brief summary in this section.

Our sample was originally selected with the objective to  understand the role galaxy-scale winds play in star formation quenching for massive, intermediate redshift galaxies. The parent sample of galaxies we use in this work is drawn from the Eighth Data Release of SDSS \citep{york00SDSS,aiha10sdssDR8}. We set out to select sources that were targeted as quasars (flagged either as \textsc{qso\textunderscore {hiz}}, \textsc{qso\textunderscore  cap}, \textsc{qso\textunderscore skirt}, \textsc{qso\textunderscore mag\textunderscore outlier}), since the SDSS QSO sample extends to fainter magnitudes than the main galaxy sample \citep{stra02SDSSmaingalaxies}. Selecting sources that have been targeted as quasars allows our sample to consist of objects that are massive and compact. The magnitude limits ensure that our sources are massive, highly star forming, and not strongly dust attenuated and the SDSS quasar selection algorithm requires that our  sources are either unresolved or that they are resolved but satisfy more stringent color-magnitude cuts. This is described in more detail in Section \ref{ssec:QSO}.

We required that our sources were spectrally classified as galaxies with apparent $16 < i < 20$. We selected sources within $0.4 <  z < 0.9$ to ensure that the MgII $\lambda \lambda 2796,2804$ line would be shifted into the optical so we could use that as a probe of galactic winds. We also exclude sources that were classified as distant red galaxies (\textsc{legacy\textunderscore target1 != DRG}). Sources with redshift warnings and bad quality plates were also thrown away. This initial cut left us with a sample of 1198 galaxies.

We fit the SDSS spectra with a combination or simple stellar population models, similar to \citet{trem04massmetal}, and a type I quasar template. From the spectral fitting, we calculated the fraction of light attributed to the quasar model ($f_{qso}$).  We also measured nebular emission and stellar absorption line indices \citep[following][]{kauf03SDSSsfhs} for the sources in our parent sample  as well as the strength of the 4000 \AA\ break (D$_{n}(4000)$) \citep{balo99dnbreak}. Our initial aim was target post starburst galaxies (PSBs) by selecting galaxies with evidence of having gone through a starburst event within the last 1 Gyr ($(\text{H}\delta_{A} + \text{H}\gamma_{A})/2$  OR $D_n(4000) < 1.2.$), but with little ongoing star formation within the last  10 Myr ([OII] 3727 \AA\  equivalent width (EW) $> -20$ \AA). These cuts reduce our sample to 645 sources.

Lastly, our sample was limited to consisting of brighter galaxies with tighter cuts on [OII] EW and including a cut on the measured quasar fraction to further ensure that strong AGN were not included. The new cuts imposed were [OII] 3727 \AA  EW $> -15$ \AA, and $f_{qso}< 0.25$. We also require that apparent $g$ and $i$ magnitudes were brighter than $g<20$ or $i<19.1$. Although we select for weak nebular emission to eliminate starbursts, many of our sources were detected in \textit{WISE} \citep{wrig10wise}, and SED fitting through the mid-infrared shows they can have SFRs$=20-500 \ \text{M}_{\odot} \ \text{yr}^{-1}$ \citep{diam12CSstarform,perr21nirspec,davi_inprep_mgii_winds}. These cuts leave us with a sample of 121 galaxies. We take advantage of the \textit{WISE} detections for our sources and make an IR color cut of $W1-W2 < 0.8$ to further limit AGN contamination \citep{ster12WISEAGN,hick17qsoSED}. The \textit{WISE} AGN cut leaves us with a population of 115 galaxies in what we are considering to be our parent sample. We include this selection criteria in our modeling of compact starburst galaxies to estimate the amount of time our galaxies would be targeted and selected by this set of criteria. A full list of targets is given in Table \ref{tab:sample} along with their redshifts, stellar masses, and SDSS photometry.

In addition to the SDSS and \textit{WISE} data for our parent sample, we also have high-S/N ($\sim 15-30$ per pixel) spectra from the Blue Channel Spectograph on the 6.5-m MMT \citep{ange79MMT}, the Magellan Echellette (MagE; \citealt{mars08MAGE}) spectrograph on the Magellan Clay telescope, and the Low Resolution Imaging Spectrometer (LRIS; \citealt{oke95LRIS}) on the Keck I telescope for 37 of the sources in our parent sample. These observations and their data reduction are detailed in \citet{davi_inprep_mgii_winds}, but broadly these observations were done using 1" slits resulting in spectra with resolution $R\sim 600-4100$. We refer to these 37 galaxies as the MgII sample.

%Of the MgII galaxies, 37/46 are included in the larger parent sample. Of the MgII galaxies that were not included in the parent sample, we are adding 6/46 with bad SDSS DR8 $z$ measurements, 1/46 targeted as a \textsc{serendipity\textunderscore first} object in the QSO selection pipeline, 3/46 objects that were slightly too faint, and 1/46 with slightly too large of an [OII] EW. 

%The sources we are including have properties that are largely consistent with our parent sample, so we still consider them to be representative compact starburst galaxies. 

\label{sec:selection}

\section{Model construction}
The aim of this work is to constrain the importance of massive, compact starburst events in galaxy quenching at $z\sim 0.5$ by estimating the space density of these objects. Here, we do this by constructing an empirical model based on the galaxies we have in our sample and then evolving a large simulated population of compact starbursts to estimate the timescales upon which they would be targeted by our selection criteria. This process can be broken down into two steps:
\begin{enumerate}
    \item Construct a set of  template distributions of stellar population parameters and SFHs  by fitting SDSS $ugriz$ model mags and \textit{W1, W2} photometry for the 115 galaxies in our sample with a Markov Chain Monte Carlo (MCMC; \citealt{metr53metropolishastings, fore13emcee}) fitter.
    \item Use the posterior distribution of SFH parameters from step 1 to predict luminous properties of a set of mock galaxies whose SFHs are consistent with our observed sample. The luminous properties are computed using the \textsc{Flexible Stellar Population Synthesis} models (FSPS; \citealt{conr09FSPS}).
\end{enumerate}

Since our small sample of galaxies consists of sources that are unresolved in SDSS imaging, we have to make a number of assumptions about their underlying stellar populations. First, we assume that the light from our compact starburst galaxies can largely be broken down into two components: a young, simple stellar population (SSP) that formed in a single, nuclear burst, and an older component that has a star formation history representative of a massive, star forming galaxy at $z \sim 0.5$. We note that there is likely clumpy star formation occurring outside of the nuclear regions of our galaxies, but due to their extremely compact \textit{HST} morphologies it is fair to assume that the contribution of these star forming regions to the total emitted light is minimal compared to the large nuclear burst. We also assume that our galaxies will only experience one burst of nuclear star formation and will then passively evolve. Although \textit{HST} observations \citep{sell14CSoutflow} showed that many of our sources have more than one core that could trigger a starburst event, we note that these sources are still unresolved in SDSS so the burst would not be localized to a particular core. This assumption is also consistent with the single burst of star formation triggered by a merger event seen in simulations \citep[e.g.][]{sprin05mergers}. Next, we naively assume that since the nuclear burst component dominates the spectral energy distribution (SED) of the total system, that the differences observed between the galaxies in our sample can solely be attributed to differences in the properties of the nuclear starburst. This assumption is consistent with the galaxies in the MgII sample having very blue spectra and young ages as derived from spectral modeling \citep[e.g.,][]{davi_inprep_mgii_winds}.

These assumptions allow us to construct a model that utilizes FSPS to simulate the stellar populations for the nuclear starburst component as well as the older, non-burst underlying stellar population. In our modeling framework, we introduce four free parameters that are fit via an MCMC routine for each of the galaxies in our sample: the age of the burst  ($t_{age}$), the fraction of total galaxy stellar mass formed in the nuclear burst ($f_{burst}$), the optical depth for the dust around young stars formed in the nuclear burst ($\tau_{dust,1}$), and the total stellar mass of the system ($M_{*}$). We separately calculate the $ugriz$, \textit{W1}, \textit{W2}, [OII] (3727 \AA) fluxes for the nuclear burst and non-burst components and their $f_{burst}$ weighted sum to determine the SED and [OII] EW for the total simulated galaxy.   

In this section, we describe the assumptions made in the FSPS modeling of both the extended non-burst and nuclear starburst components as well as the MCMC fitting we use to constrain values for the free parameters in our model.

For both, the non-burst and nuclear burst components, we make the following assumptions. We assume a \citet{chab03IMF} initial mass function (\textsc{imf\textunderscore type = 1}) and $\log Z/Z_{\odot} = -0.3$ metallicity (\textsc{logzsol = -0.3}) using the $M_{*}-Z$ relation presented in \cite{gill21metallicitySF} calibrated for solar $12 + \log (\text{O/H}) = 8.66$ and $Z_{\odot} = 0.0121$. We set \textsc{add\textunderscore neb\textunderscore emission = true} to allow for nebular emission from \textsc{CLOUDY} models \citep{byle17cloudyFSPS}. We assume \citet{char00extinction} extinction (\textsc{dust\textunderscore type = 0}) with \textsc{dust\textunderscore tesc = 7} ($\log(t_{age}/ \text{yr})$) \citep[e.g][]{blit80GMClifetime,char00extinction,conr09FSPS}, where  \textsc{dust\textunderscore tesc} is the age in \citet{char00extinction} extinction model at which stars are attenuated by $\tau_{dust,1}$ and $\tau_{dust,2}$.  We also set \textsc{agb\textunderscore dust = true} since IR SEDs of star forming galaxies are poorly fit without incorporating dust shells around AGB stars \citep{vill18AGBdust}.

\subsection{Modeling the extended, non-burst  component}
\begin{figure*}
\centering
\includegraphics[width=\textwidth]{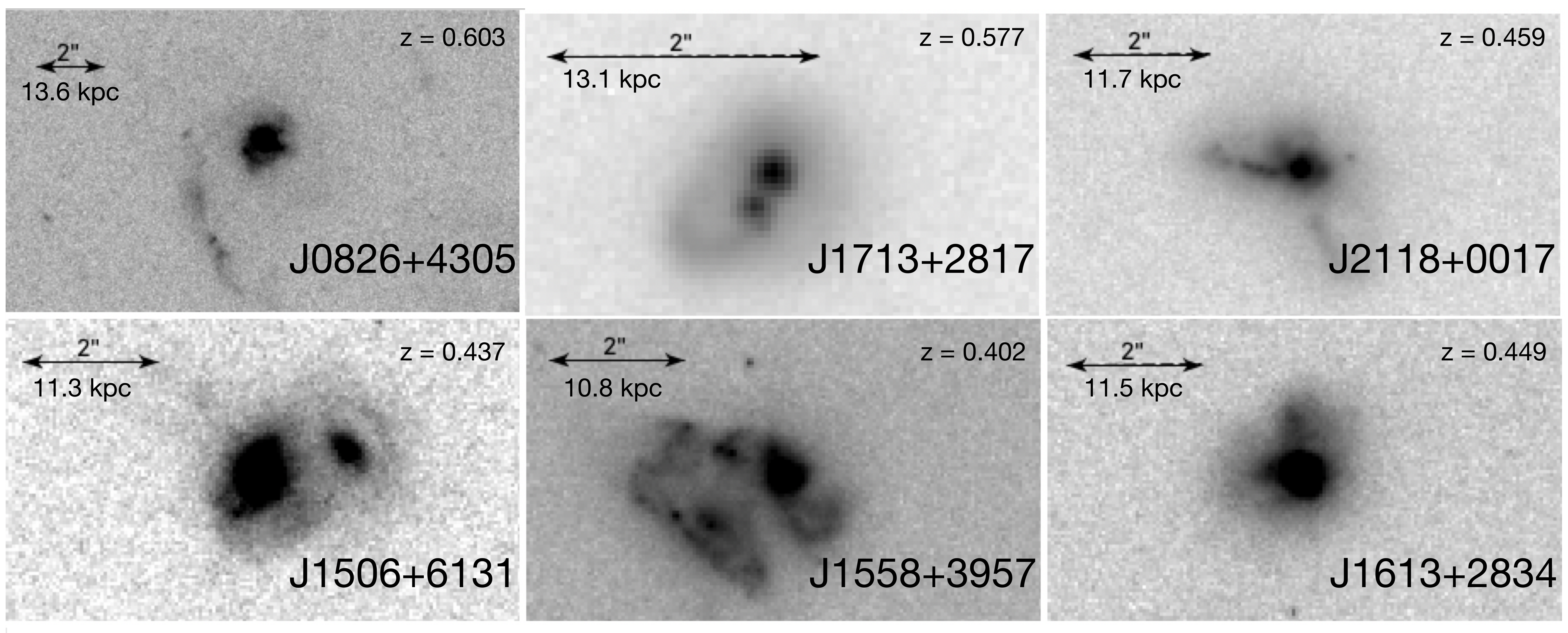}
\caption{\textit{HST} WFC3 cutouts of 6 representative galaxies in our sample that overlap with those presented in \citet{sell14CSoutflow}. We note that we omit J0944+0930 and J1104+5946 from \citet{sell14CSoutflow} as they do not satisfy all of our selection criteria. All of these galaxies show clear signs of tidal disruptions, consistent with their extreme nuclear starbursts being triggered by major merger events.}
\label{fig:cutouts}
\end{figure*}

The photometric and morphological properties of the extended stellar population are most important in the later stages of the compact starburst's evolution since the contribution of the nuclear burst wanes over time. Here, we describe the assumptions we make in the FSPS modeling of the extended, non-burst component. We initialize FSPS such that \textsc{tage} is the Hubble time (in Gyr) at the redshift of a given galaxy, \textsc{dust1 = 1}, and \textsc{dust2 = 0.5}. %We chose an age of 6 Gyr to ensure that the stellar population would be younger than the Hubble time for the highest-$z$ galaxies in our sample to reduce computational overhead. 
%We chose these dust optical depths so they would be less than the median \citet{char00extinction} values since we will be adding additional dust in the nuclear burst. 
We chose these dust optical depths to ensure that the $ug$ photometry for the modeled extended stellar component would be fainter than that of the reddest observed sources in our sample, while being consistent with the recommended values given in \citet{char00extinction}. We explored the effects of changing \textsc{tage} and the dust parameters for the extended components in the galaxies shown in Figure \ref{fig:cutouts} to ensure that our modeling is largely robust to extended component assumptions and found that the results of our MCMC fitting do not change with changing non-burst initial conditions.

A crucial piece to modeling the stellar population of the extended, non-burst component is assuming a particular star formation history (SFH). \textit{HST} images show hints of a smooth, extended underlying stellar population \citep{diam21CSoutflows}. The presence of tidal features in our \textit{HST} observations suggests that the galaxies in our sample have recently undergone merger events, and their high star formation surface densities indicate that that these mergers were likely gas rich \citep[e.g.,][]{diam12CSstarform,sell14CSoutflow}. Based on this, we assume that the extended, non-burst stellar populations have a star formation history typical of actively star forming disk galaxies.

However, the SFHs of star forming disk galaxies are uncertain. There are many possible SFHs that would be able to build up the tightly-correlated star formation main sequence at late cosmic times \citep[e.g.][]{oeml17SFHdisks}. For simplicity, since young stars dominate the light output from a stellar population we approximate the SFH as being flat over cosmic time to ensure that the progenitor galaxies in the system were experiencing some degree of star formation prior to merging. We do this by setting the FSPS SFH parameter as a delayed-burst SFH (sfh = 4 in FSPS) but with the constant star formation fraction set to 1.

We also note that we explored other SFHs that peaked at earlier cosmic times, such as the dark matter halo mass dependent models constructed in \citet{behr19UMAchine}, but our MCMC chains for these models were not able to reach convergence. The inability for our chains to converge is consistent with the fact that we do not believe that \citet{behr19UMAchine}-like SFHs would be  physically representative of galaxies like those in our sample. For massive ($M_{*} \sim 10^{11} \ \text{M}_{\odot}$) galaxies like the ones in our sample, this would suggest that our sources would have peaked in star formation at $z\sim 2$ and then passively evolved until $z\sim0.5$. This would imply that the progenitors of our compact starbursts would be almost entirely be quiescent, which is unlikely do to their high gas fractions. Therefore, we do not include models like this in our analysis.

\subsection{Modeling the nuclear burst}
Recent observational evidence has shown that intermediate redshift, extreme compact starburst galaxies are likely to exhibit flat age gradients, meaning that their optical light is dominated by star formation that began and ended in one uniform event \citep[e.g.,][]{sett20cs_ages}. Since we expect all of the stars formed in the nuclear burst to have formed at approximately the same time, we model the starburst as a simple stellar population (SSP) in FSPS (sfh = 0). This choice is consistent with very short burst durations we derive from non-parametric SFH modeling of a subset of our sample with high S/N spectra \citep{geac18molecoutflow, trem_inprep_sample, davi_inprep_mgii_winds}. This work (detailed in \citet{davi_inprep_mgii_winds}) is done by fitting the rest frame UV-mid IR broadband photometry and high-resolution spectra simultaneously using Prospector \citep{leja19prospector,john21prospector}. We also assume that the dust in the vicinity of the nuclear starburst extincts some of the light from the newly formed stars. We leave the age of the central burst ($\log t_{age}$) and the optical depth ($\tau _{burst}$) as free parameters that will later be constrained with MCMC fits to the photometric data of the sources in our observed sample. We set \textsc{dust2} = $\tau _{burst}/2$ \citep[e.g.,][]{wild11attenuation}. We similarly calculate SDSS $ugriz$ and \textit{WISE} W1 \& W2 magnitudes for the nuclear bursts as we did for the extended, non-burst stellar population.

\subsection{Calculating PSF magnitudes}
Once we have the model photometry for the extended, non-burst stellar populations and their nuclear bursts, we can combine them to get the photometry for the entire system. We start by converting the modeled apparent AB magnitudes for the extended, non-burst stellar population and the burst component to flux densities.
%\begin{equation}
%$$f_{\nu} = 10^{\frac{m_{ab} + 48.6}{-2.5}} \ \text{erg} \ \text{s}^{-1} \ \text{cm}^{-2} \ \text{Hz}^{-1}.$$
%ation}
The output magnitudes of FSPS are normalized to $1 \ \text{M}_{\odot}$ at every epoch, so we calculate the fluxes for our galaxies and nuclei by multiplying their $1 \ \text{M}_{\odot}$ flux densities by their respective masses. We define the mass of the nuclear burst as $M_{nuc} = f_{burst}  \times M_{*}$ and $M_{host} = (1 - f_{burst}) \times M_{*}$. We also leave $f_{burst}$ and $M_{*}$ as free parameters in our MCMC fitting in addition to $\tau _{dust}$ and $\log t_{age}$ as described earlier.

For sources observed in SDSS, the QSO targeting pipeline takes a source's $ugriz$ PSF magnitudes as input rather than its de Vaucouleurs or exponential disk model magnitudes \citep{rich02SDSSquasar}. The output magnitudes from FSPS are representative of model magnitudes, so we must first convert these to PSF magnitudes before we run the SDSS QSO targeting algorithm on our modeled sample. We do this by first assigning surface brightness profiles to both components of the galaxy. For the extended, non-burst component, we assume a $n=1$ S\'{e}rsic profile where the effective radius ($R_{\text{eff}}$) is taken from the redshift-dependent star forming galaxy size-mass relation presented in \citet{mowl19sizemass}. Due to the nuclear starburst's compact nature, we assume a $n=4$ S\'{e}rsic profile where $R_{\text{eff}}$ is $\sim 300$ pc, as motivated by observations \citep[e.g.,][]{geac13CSmolec,sell14CSoutflow}. \citet{diam21CSoutflows} showed that $R_{eff} < 1$ kpc  for the \textit{HST}-observed galaxies. We do not vary $R_{eff}$ for the nuclear components for our modeled galaxies since $\sim100$ pc scale starbursts would always be unresolved in SDSS and are effectively observed as point sources.

We convert $R_{\text{eff}}$ for each component from kpc to arcsec using their cosmological angular size distances and normalize the surface brightness profiles ($I(r)$) for each component such that $$2 \pi \int ^{\infty} _{0} I_{comp}(r) r dr = f_{\nu , comp}.$$ We then convolve these component surface brightness profiles with the SDSS PSF in each photometric band. The full width half maxes (FWHMs) for the $ugriz$ bands are 1.53, 1.44, 1.32, 1.26, 1.29 arcsec, respectively. The convolved burst and disk components are then added together to create a modeled total galaxy surface brightness profile. We then fit this profile with a 2D-Gaussian model of the SDSS PSF and integrate the Gaussian model fit to obtain PSF fluxes in each respective band. The PSF fluxes are then converted to apparent AB magnitudes so they could later (\S4.1) be passed through the SDSS QSO selection pipeline.

%To maintain consistency with the SDSS QSO quasar targeting algorithim, we calculate PSF magnitudes for our modeled galaxies by fitting surface brightness profiles to each component of the galaxy. For the extended, non-burst component, we assume a $n=1$ S\'{e}rsic profile where the half-light radius ($R_{\text{eff}}$) is taken from the redshift-dependent star forming galaxy size-mass relation presented in \citet{mowl19sizemass}. Due to the nuclear starburst's compact nature, we assume a $n=4$ S\'{e}rsic profile where $R_{\text{eff}}$ is on the order of hundreds of parsecs, as motivated by observations \citep[e.g.,][]{trem07CSoutflow,sell14CSoutflow}. We convert $R_{\text{eff}}$ for each component to from kpc to arcsec using their cosmological angular size distances so that all of our values are consistent with the sizes that would be directly measured from SDSS observations. 
%ation}
%We then integrate the surface brightness profile for each SDSS photometric band on the interval $0 < r < R_{seeing}$, where $R_{seeing}$ is the seeing in arcsec for each band, the values of which can be found in Table \ref{tab:seeing}. This gives us the fraction of light contained within the PSF aperture for each filter. We then multiply this light fraction by the total flux density to obtain the PSF flux for each component. We then add these flux components together and calculate absolute magnitudes to obtain the photometry for the simulated compact starburst galaxy.

\subsection{Constraining model free parameters with MCMC}

We have constructed a 4-parameter model for the photometry and [OII] (3727 \AA) EW of intermediate-$z$ compact starbursts by utilizing FSPS. FSPS directly outputs model mags and spectra of stellar populations. We calculate [OII] (3727 \AA) EW from the FSPS output spectrum using \textsc{specutils} \citep{specutils}. As stated above, our compact starburst model is the sum of separately modeling the host galaxy and nuclear burst contributions to the overall photometric and spectral properties. In this model, we leave the age of the nuclear starburst ($\log \ t_{age}/\text{Myr}$), the burst fraction ($f_{burst}$), optical depth of dust extincting young stellar light ($\tau _{dust}$), and the galaxy stellar mass ($\log \ M_{*}/ M_{\odot}$) as free parameters. Here we detail how we constrain possible parameter values using MCMC fitting to the $ugriz$ and W1/W2 photometry for our observed galaxies.

\subsubsection{Parameter fitting}
As discussed in Section \ref{sec:selection}, our collaboration has been studying a sample of 115 intermediate-$z$ compact starburst galaxies. Archival SDSS $ugriz$ and \textit{WISE} W1 and W2 photometry are available for the full parent sample. For each of these, we constrain the probability densities for $\log \ t_{age}$, $f_{burst}$, $\tau _{dust}$, and $\log M_{*}$ using the ensemble adaptation of the Metropolis-Hastings MCMC algorithm from the package, \textsc{emcee}  \citep{metr53metropolishastings,fore13emcee}. Each step of our MCMC calculates the model SDSS $ugriz$,  \textit{WISE} W1, and W2 photometry, and compares them to those for each observed galaxy. For each galaxy, we run the MCMC such that the autocorrelation time for each walker is $\sim 50$ times less than the run time. For most of our galaxies this is $\sim 60,000$ steps. We use the \textsc{emcee} ensemble stretch move with scale parameter $a=2$. We randomly initialize each walker in the intervals
$$0.5 < \log \ t_{age}/\text{Myr} < 2$$ $$0.05 < f_{burst} < 0.4$$ $$0.3 < \tau _{dust} < 1$$  $$10 < \log M_{*}/M_{\odot} <11$$ and allow them to explore the parameter space $$0.5 < \log \ t_{age}/\text{Myr} < 3$$ $$0.05 < f_{burst} < 0.65$$ $$0 < \tau _{dust} < 5$$ $$10 < \log M_{*}/M_{\odot} <12$$ such that it finds the parameter values that are most likely to minimize the difference between the model and observed photometry.

\begin{figure*}
\centering
\includegraphics[width=0.8\linewidth]{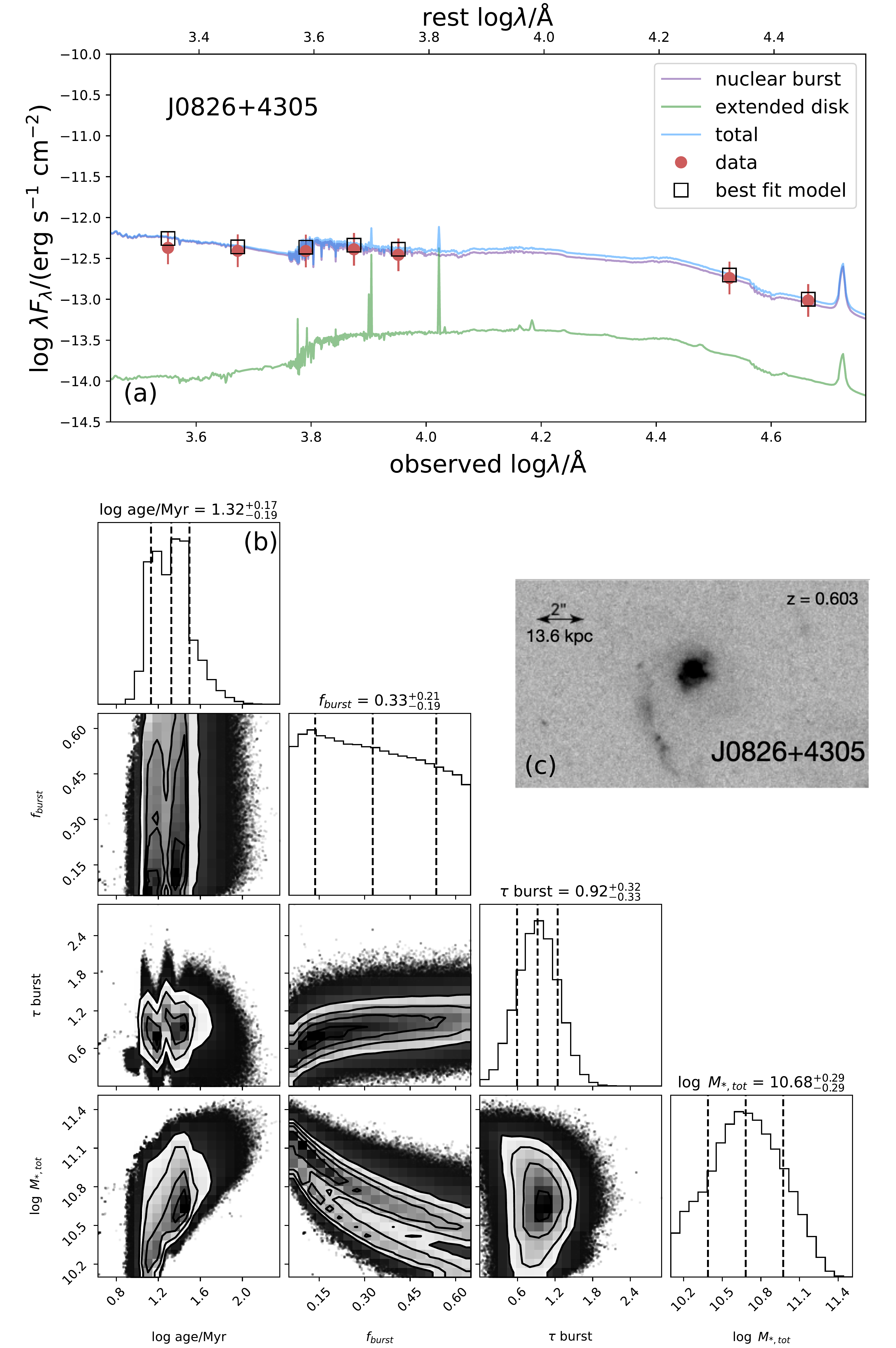}
\caption{\textit{Panel (a)}: Best fit SED for galaxy J0826+4305. The red points and error bars are the observed photometry and $\pm 0.25$ magnitude uncertainty region, respectively. The open black squares are the modeled photometry. The blue, violet, and green curves are the modeled SED for the total galaxy system, nuclear burst, and host galaxy, respectively. \textit{Panel (b)}: Triangle plot of parameter posterior distributions for galaxy J0826+4305. We calculate the mean and covariances of these posterior distributions to model them as 4D-Gaussian distributions. We then randomly draw sets of parameter values from the Gaussian-modeled posterior to construct a mock population of compact starbursts. \textit{Panel (c)}: Galaxy cutout as seen in Figure \ref{fig:cutouts}.}
\label{fig:mcmc}
\end{figure*}

For each galaxy in our sample, we output the mean parameter values and their covariance from MCMC-calculated posterior distributions. We use these mean values and their covariances to model these posteriors as 4-dimensional Gaussian distributions whose means and standard deviations are identical that of the MCMC output. We do this to reduce noise later in our analysis since we use these distributions to randomly draw sets of parameter values to model mock galaxies based on the ones in our observed sample. The best fit SED and parameter probability distributions for a constantly star forming host based on the galaxy J0826+4305 can be seen in panels (a) and (b) Figure \ref{fig:mcmc}, respectively.We also include these for J1713+2817, J2118+0017, J1506+6131, J1558+3957, and J1613+2834 in Figures \ref{fig:j1713_app}, \ref{fig:j2118_app}, \ref{fig:j1506_app}, \ref{fig:j1558_app}, and \ref{fig:j1613_app}, respectively. For consistency with other studies of our objects, we note general agreement between our best fit stellar masses and those presented in \citet{sell14CSoutflow} for the galaxies that were included in both of our samples. This is shown in Table \ref{tab:sell_compare}.

%We model the resulting probability densities as 4-dimensional Gaussian distributions whose covariance matrices are taken from the output of the MCMC. The best fit SED and parameter probability distributions for a constantly star forming host based on the galaxy J0826+4305 can be seen in panels (a) and (b) Figure \ref{fig:mcmc}, respectively.

For each of the 115 galaxies in our sample we randomly draw $\log \ t_{age}$, $f_{burst}$, $\tau _{dust}$, and $\log M_{*}$ values from their respective Gaussian-modeled posterior distributions taking into account the covariances between each of the parameters, to model a population of galaxies with properties similar to the observed source. We can then evolve these modeled galaxies to estimate a distribution of selectable lifetimes for each of the galaxies in our sample.

%From here, can we randomly draw $\log \ t_{age}$, $f_{burst}$, $\tau _{dust}$, and $\log M_{*}$ from the constant SFH host model posterior distributions to model a large population of compact starbursts based on the observed galaxies in our sample.

\section{Modeling the targeting algorithm \& selection function }
The ultimate goal for our model is to be able to estimate the space density of $z\sim0.5$, massive, compact starburst galaxies. To do this, we need to understand the timescales upon which these galaxies would be selected under a set of targeting criteria. Here, we detail how we model the various components of the selection function we use to identify sources in our sample.

\subsection{The SDSS QSO targeting algorithm}
\label{ssec:QSO}
All of the sources in our observed sample were initially targeted for SDSS spectroscopy as QSOs based their bright magnitudes and blue colors. In order to ensure that our modeled galaxies would satisfy these criteria, we need to incorporate this selection into our modeled targeting function.

The SDSS QSO targeting algorithm identifies sources based on their location in three-dimensional color space. This is the $(u-g)$-$(g-r)$-$(r-i)$ ($ugri$) color cube for $z<3$ sources and $(g-r)$-$(r-i)$-$(i-z)$ ($griz$) cube for galaxies at higher redshifts. The QSO catalog constructed from SDSS DR8 sources was selected using the \citet{rich02SDSSquasar} targeting algorithm \footnote{Python adaptation of \citet{rich02SDSSquasar} QSO selection algorithm can be found at \url{www.github.com/ke27whal/sdss\_qso\_selection}.}. The SDSS quasar selection function aims to identify sources that lie far from the region of color space where stars are most likely to be found as well as for sources to satisfy general color/magnitude cuts. All magnitudes referenced in the targeting algorithm are PSF magnitudes. Since we are working with modeled data that is free from observational uncertainty, we do not include the steps in the algorithm that flag sources for having data with fatal errors. 

Since quasars and local stars both exhibit bright apparent magnitudes and are unresolved point sources, the algorithm needs to be able to differentiate between them in color-color-color space. The algorithm makes use of the method described in \citet{newb97locus} that defines a ``stellar locus'' in color-color-color space where stars are most likely to exist. The stellar locus is constructed by analyzing the distribution of SDSS identified stars in color space. To maintain generality, we will refer to the main coordinate system describing the color-color-color cube as $\langle \hat{x}, \hat{y}, \hat{z \rangle}$, where $\hat{x}$ is in the direction of the bluest color axis and $\hat{z}$ in the direction of the reddest. The locus construction algorithm begins by setting the endpoints of the stellar distribution in color space and then iteratively calculating midpoints. This process allows a local coordinate system ($\langle \hat{i}_{i}, \hat{j}_{i}, \hat{k}_{i} \rangle$) to be defined at each locus point. At each locus point ($p_{i}$), $\ \hat{k}_{i}$ is defined as a unit vector in the direction $\overrightarrow{p_{i+2} -p_{i}}$. As detailed in \citet{newb97locus}, unit vectors $\hat{i}_{i}$, $\hat{j}_{i}$, and $\hat{k}_{i}$ are given as
$$ \hat{k}_{i} \equiv k_{x} \hat{x} + k_{y} \hat{y} + k_{z} \hat{z}, $$ $$\hat{j}_{i} \equiv (\hat{k}_{i} \times \hat{z})/|\hat{k}_{i} \times \hat{z} | = (k_{y} \hat{x} - k_{x} \hat{y})/\sqrt{k_{x}^{2} + k_{ y}^{2}},$$ $$\hat{i}_{i} \equiv \hat{j}_{i} \times \hat{k}_{i} = [-k_{x}k_{z} \hat{x} - k_{y}k_{z} \hat{y} + (k_{x}^{2} + k_{y}^{2})\hat{z}]/\sqrt{k_{x}^2 + k_{y}^{2}}.$$ 

The cross section of the stellar locus is measured by fitting an ellipse perpendicular to $\hat{k}_{i}$ at each point. The semi-major and semi-minor axes of the ellipses are in the direction of unit vectors $\hat{l}_{i}$ and $\hat{m}_{i}$, respectively, and are defined as 
$$\hat{l}_{i} \equiv \hat{i}_{i} \cos \theta_{i} + \hat{j}_{i} \sin \theta _{i},$$ $$\hat{m}_{i} \equiv - \hat{i}_{i} \sin \theta _{i} + \hat{j}_{i} \cos \theta _{i}$$ where $\theta _{i}$ is the angle  between the major axis of the ellipse and unit vector $\hat{i}$. We adopted the locus point positions, $\theta_{i}$, $\hat{k}_{i}$, $| \vec{l}_{i} |$, and $| \vec{m}_{i} |$ values from \citet{rich02SDSSquasar}, and proceeded to construct right cylinders that define the $4 \sigma$ stellar locus probability region in color-color-color space. We also incorporate the mid-$z$ inclusion region as the white dwarf/A star exclusion regions detailed in \citet{rich02SDSSquasar}.

Sources targeted as quasars must also satisfy color and magnitude cuts in addition to not belonging to the stellar locus. For low-$z$ sources in the \ugri  color cube, all objects must have apparent $i$-band magnitude $15 < i < 19.1$ \citep{rich02SDSSquasar}. Both extended and point source objects are allowed to be selected as quasars, but they need to satisfy different sets of criteria. Point source objects only need to fulfill the magnitude and stellar locus cuts to be targeted. Extended sources are kept if they are likely to contain an active nucleus. This is most likely when $(u-g) < 0.9$, as redder AGN would be at high-$z$ and would not be extended \citep{rich02SDSSquasar,adel06dr4}. This $(u-g)$ cut does not remove blue, extended star forming galaxies, so a second cut of $l_{i} > 0$ and $m_{i} >0$ is applied where $l_{i} $ and $m_{i}$ are positions within the $\langle \hat{k}, \hat{l}, \hat{m} \rangle$ coordinate space defined earlier. In the high-$z$ \griz color cube, all outliers from the stellar locus with $15 < i < 20.4$ are targeted as quasars. However, to avoid contamination from low-$z$ quasars, sources are removed from the high-$z$ sample when all of the following criteria are met; $$(g-r) < 1.0,$$ $$(u-g) \geq 0.8,$$ $$i \geq 19.1 \ \ \text{OR} \ \ (u-g) < 2.5.$$

We allow the sources in our sample to be targeted as either low-$z$ or high-$z$ quasars since our observed sample contains a mixture of both target types.

\subsection{Spectroscopic/photometric selection}
In addition to being blue, unresolved sources, the galaxies in our sample also exhibit weak nebular emission characteristic of post starburst galaxies. As mentioned earlier, we implement an emission line equivalent width (EW) cut on  [OII] (3727 \AA) such that [OII] EW$> -15$ \AA, consistent with that used for our parent sample \citep{sell14CSoutflow,davi_inprep_mgii_winds,trem_inprep_sample}. We also model the $g < 20$ flux limit and $W1-W2 < 0.8$ \textit{WISE} color cut that we impose on our sample. 

\section{Estimating the space density}
In this section, we discuss the various parameters that contribute to the calculated compact starburst space density ($n_{CS}$) as well as the possible sources of uncertainty. We estimate the space density in the redshift range $0.4 <  z < 0.9$ as 
\begin{equation}
    n_{CS} \sim \frac{N_{targeted}}{f_{complete}} \cdot \frac{t_{cosmic}}{V_{0.4 <  z < 0.9}} \cdot \frac{A_{sky}}{A_{SDSS}} \cdot \biggl< \frac{1}{ t_{obs}}\biggr>.
\label{eq:space_density}
\end{equation} 

Here, $N_{targeted}$ is defined as the number of galaxies in our observed sample of massive, compact starburst galaxies, $f_{complete}$ is the completeness of the SDSS QSO catalog ($f_{complete}  \sim 0.9$; \citealt{vand05QSOcompleteness}), $V_{0.4 <  z < 0.9} \ $ is the volume in Mpc$^{-3}$ contained within the redshift range $0.4 <  z < 0.9$, $A_{SDSS}/A_{sky}$ is the fractional area of the SDSS footprint relative to the area of the entire sky, $t_{cosmic}$ is the amount of cosmic time in Myr contained in the redshift range $0.4 <  z < 0.9$, and $\langle 1/t_{obs} \rangle $ is the average of the inverse selectability timescale in Myr. The only model-dependent factor in this calculation is the amount of time our sources would be selected under a particular set of targeting criteria, so we will spend the first part of this section focusing on calculating this value.

It is also worth highlighting that the timescale we are calculating for our sources is the amount of time these objects would be targeted under our set of selection criteria. This is a separate quantity from the amount of physical time galaxies might be undergoing an extremely compact starburst phase. The physical timescale is also dependent on how we define these sources. A unifying feature of the observed sources in our sample is that they are late-stage major mergers that host extremely young stellar populations. It is possible that some of them have quenched/are very recent PSBs and that others are still forming stars. Broadly, we define our sources as galaxies that have recently experienced an extreme nuclear burst of star formation. Calculating the physical timescale for these sources would require much more detailed modeling which is beyond the scope of this work. Our goal here is to estimate the space density of objects that would be targeted by our selection criteria at some point in their evolution.

\subsection{Calculating observed lifetimes}

\begin{figure*}
    \centering
    \includegraphics[width=1\textwidth]{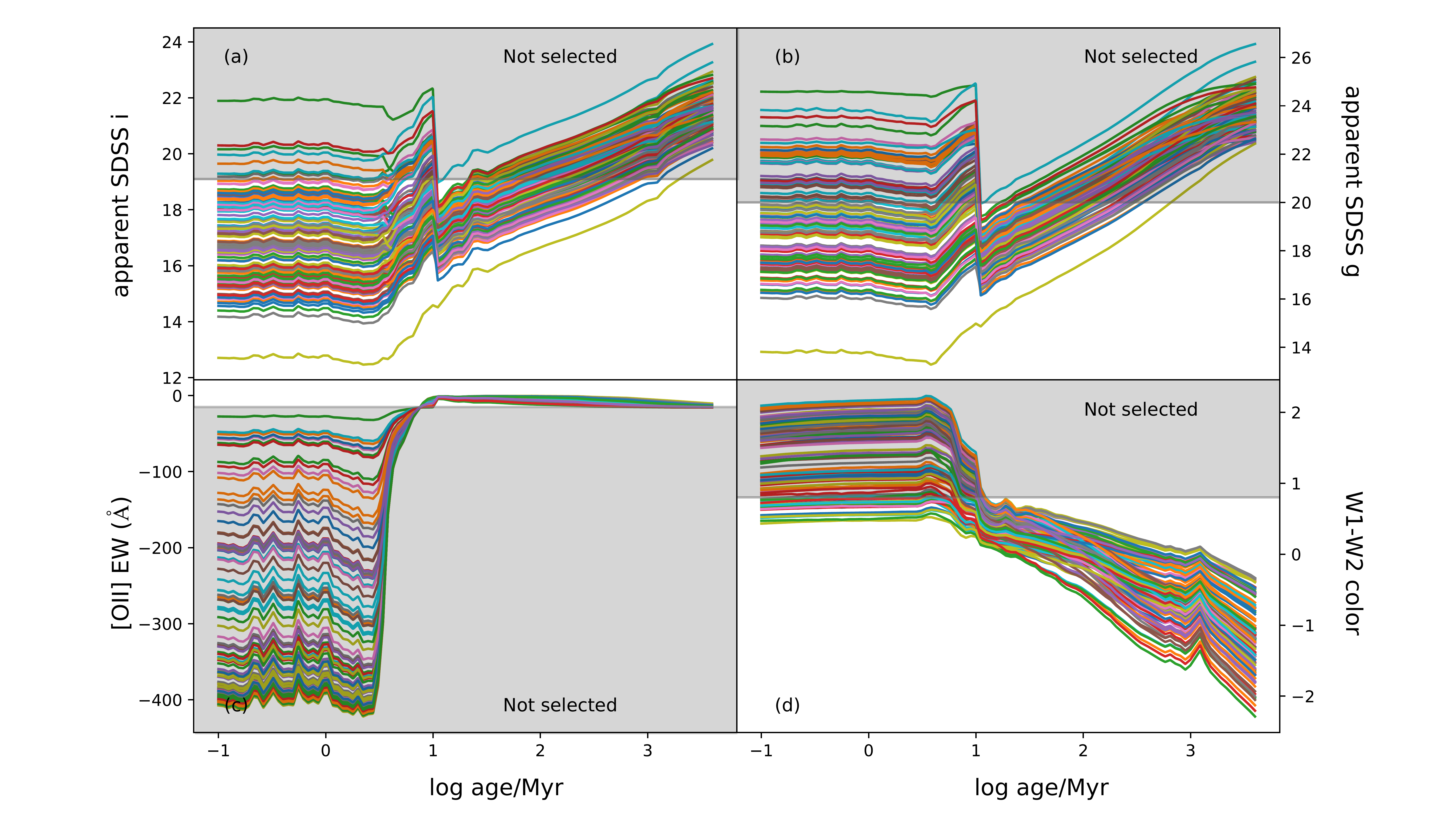}
    \caption{Shown here are the modeled evolutionary tracks of the apparent $i$-band and $g$-band \textit{SDSS} magnitudes (\textit{panels (a) \& (b))}, [OII] equivalent width (\textit{panel (c)}), and \textit{WISE} $W1-W2$ color (\textit{panel (d)}) for a sub-sample of modeled galaxies. The x-axis is age relative to the burst peak. The grey-shaded rectangles represent the regions of parameter space that would not be selected by the criteria placed on that given parameter. This is a schematic representation--- the full details of our source selection can be found in Section \ref{sec:selection}.}
    \label{fig:evolution}
\end{figure*}

For each of the 115 galaxies in our sample, we used SDSS $ugriz$ model mags and \textit{WISE} W1/W2 measured photometry to construct SEDs which were then fit by our MCMC routine to obtain the posterior distributions  for $\log t_{age}/\text{Myr}$, $f_{burst}$, $\tau_{dust}$, and $\log \ M_{*,tot}/M_{\odot}$. These posterior distributions were then modeled as 4-dimensional Gaussian distributions and we output their covariance matrices. For each of the 115 observed galaxies in our sample, we draw 200 sets of parameters from the respective posterior distributions while taking into account covariances between parameters. This gives us 115$\times$200 mock galaxies which we then evolve. We evolve our modeled galaxies within the time interval $-1 < \log \ t_{age}/{\text{Myr}} < 2.5$ in 1000 uniformly spaced steps. We calculate [OII] EWs from the output FSPS spectrum using \textsc{specutils} \citep{specutils}, as well as the photometry at each step to determine if the sources would be targeted by our selection criteria at each time step. This allows us to construct selected lifetime distributions for each of the 115 observed galaxies in our sample. The evolutionary tracks for a subset of randomly selected galaxies' $i$ and $g$-band magnitudes, [OII] EWs, and $W1-W2$ colors, as well as the selection limits on each respective parameter can be seen in Figure \ref{fig:evolution}. We note that Figure \ref{fig:evolution} does not include the SDSS QSO targeting selection since that is a much more complicated set of criteria and would be impossible to visually display. However, we do apply it in our target selection.

\begin{figure}[h]
    \centering
    \includegraphics[width=0.5\textwidth]{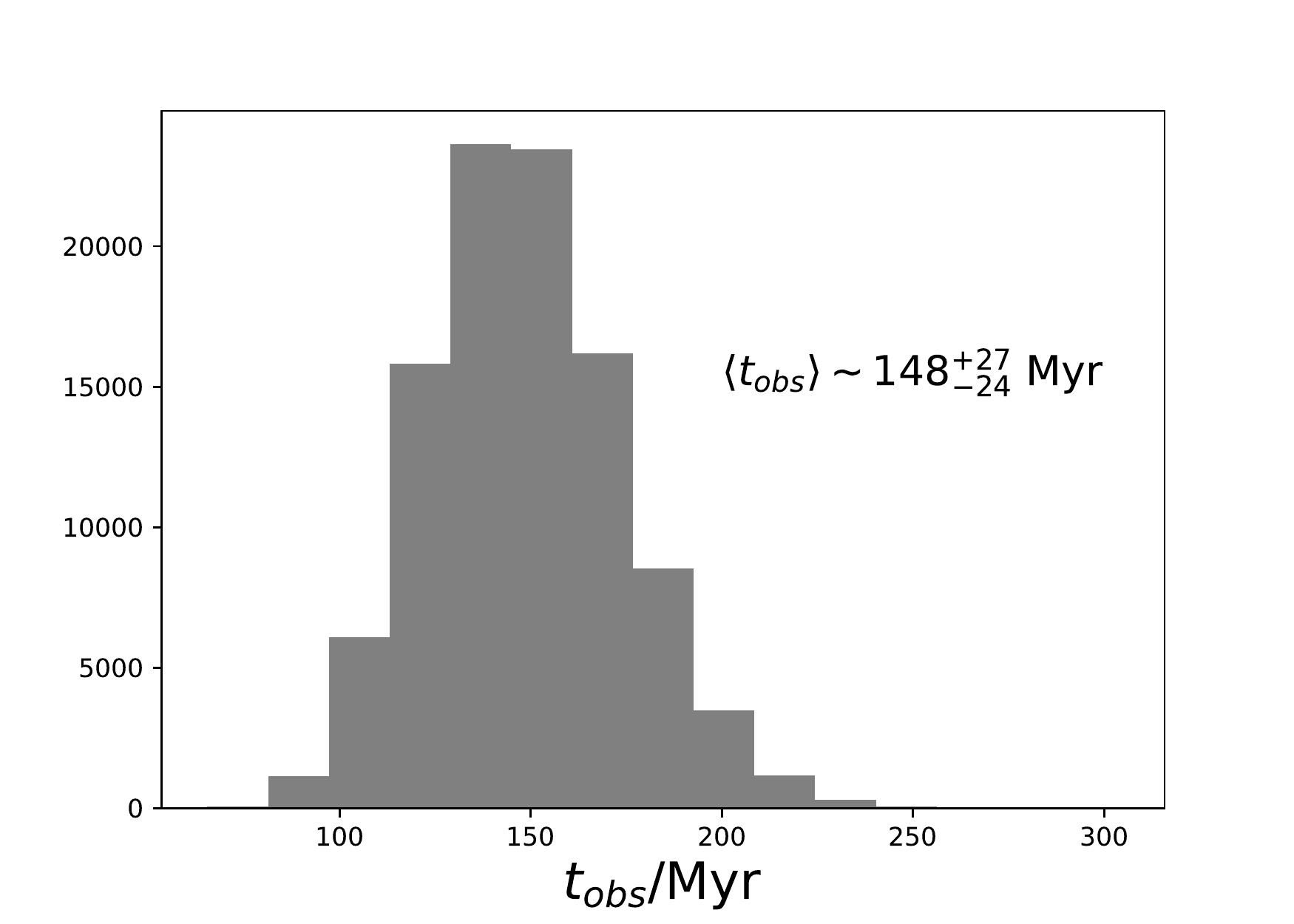}
    \caption{Distribution of average selected lifetimes from the mock sample. We find that extreme nuclear starbursts like the ones observed in our galaxies would be selected for $\sim 148 ^{+27}_{-24}$ Myr, consistent with the burst ages calculated in \citet{davi_inprep_mgii_winds}.} 
    \label{fig:obs_lifetime}
\end{figure}

In the following section, we detail how we determine the space density of our sources by randomly sampling with replacement the selected lifetime distribution calculated by evolving mock galaxies. In short, we bootstrap by generating 100,000 randomly sampled (with replacement) populations of 115 mock galaxies. For each iteration, we randomly draw an array of 115 indices which correlates to the various observed galaxies in our sample. We use the randomly drawn indices to pull selected lifetimes from the corresponding selected lifetime distributions. We then average these lifetimes to determine a selectability timescale for that given mock population of galaxies. The average selected lifetime distribution for the 100,000 samples of 115 mock galaxies is shown in Figure \ref{fig:obs_lifetime}. We find that on average, compact starburst galaxies like the ones we observe would be selected under our set of targeting criteria for $148 ^{+27}_{-24}$  Myr. This timescale is broadly consistent with the average post-starburst peak age of $70\pm106$ Myr calculated in \citet{davi_inprep_mgii_winds}. 

In our modeling, we find that our mock galaxies would be targeted soon after the nuclear burst occurs, meaning that we can directly compare our selectability timescale and the post-starburst peak SF ages in \citet{davi_inprep_mgii_winds}. The light-weighted stellar ages of the MgII sample ranging from $\sim$ 13-300 Myr) galaxies are consistent with the calculated selectability timescale in this work. This is a good consistency check to ensure that our modeling shows that galaxies in our observed sample would be selectable at their best-fit stellar ages.

%In the following section, we detail how we determine the space density of our sources by randomly sampling with replacement the observed lifetime distribution calculated by evolving mock galaxies. We iterate this part of the calculation 100,000 times to determine a statistical mean lifetime and standard deviation for a large mock sample of galaxies. In Figure \ref{fig:obs_lifetime}, we show the  average distribution of randomly selected lifetimes for all of the galaxies in our mock sample, as determined by re-iterating through averaging the lifetimes drawn from a random sample with replacement of the observable lifetime distribution. 

We next use the selectability timescales of our modeled compact starburst galaxies to estimate their space density.
%Along with light-weighted ages, \citet{davi_inprep_mgii_winds} presented SFHs for 3/46 MgII galaxies in our sample. These SFHs show that these burst events also occur on $\sim 100$ Myr timescales. 

\subsection{Calculating space density}
\label{sec:spacedensity}

As stated above, we estimate the space density in the redshift range $0.4 <  z < 0.9$ \  (Equation \ref{eq:space_density}) by randomly sampling from our selected lifetime distributions. To ensure that we sample a sufficiently large population of mock galaxies, we iterate this part of the calculation 100,000 times. 

For each  of the 100,000 iterations, we randomly sample with replacement 115 galaxies from our mock sample. For each of the galaxies in that sample, we randomly draw a $\log t_{obs}/\text{Myr}$ value from the observable lifetime distribution that corresponds to that particular galaxy. In each iteration, we use these $\log t_{obs}/\text{Myr}$ values to compute,  \begin{equation}
\biggl< \frac{1}{ t_{obs}}\biggr> = \frac{1}{N_{sim}}\sum_{i}^{N_{sim}} \ \bigg(\frac{1}{t_{obs,i}} \bigg),\end{equation} 
where $N_{sim} = 115$. We then use this to calculate the space density for the random population generated each iteration using the expression above. \begin{figure}[h]
    \centering
    \includegraphics[width=0.5\textwidth]{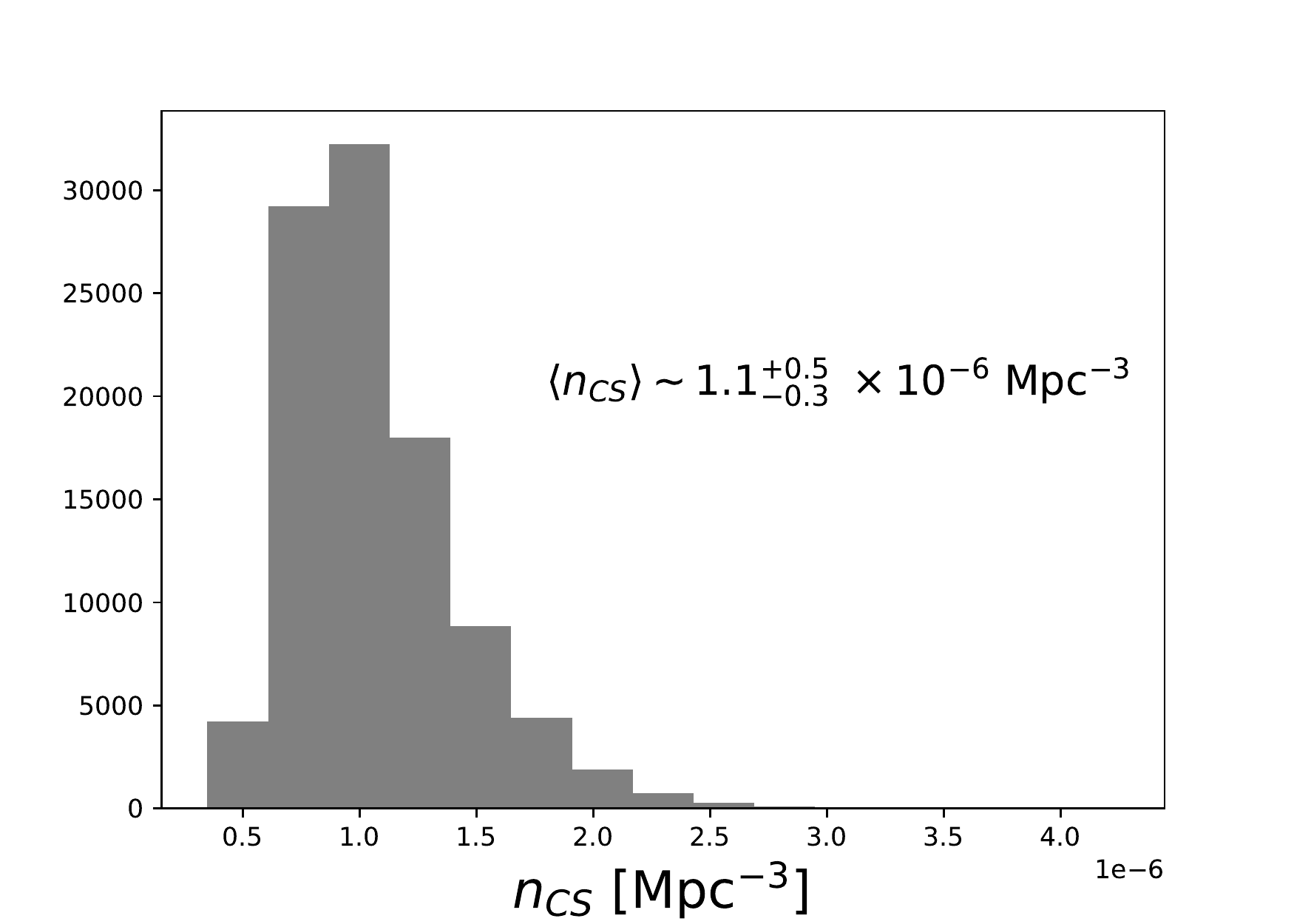}
    \caption{Space density distribution calculated from our mock population of galaxies. We estimate that the space density for our population of $0.4 <  z < 0.9$ compact starburst galaxies is $(1.1 ^{+0.5}_{-0.3}) \times 10^{-6} \ \text{Mpc}^{-3}$.} 
    \label{fig:space_density}
\end{figure} The resulting space density distribution (calculated using  Equation \ref{eq:space_density}) can be seen in Figure \ref{fig:space_density}. We estimate the space density of these massive, compact starbursts to be $(1.1 ^{+0.5}_{-0.3}) \times 10^{-6} \ \text{Mpc}^{-3}$ in the redshift range $0.4<z<0.9$.

\section{Cosmological context}
  One of the most interesting questions surrounding our sample of galaxies is whether or not this type of compact starburst phase is characteristic in the evolution of many, if not most, massive galaxies. A widely supported view of galaxy formation and evolution is that mergers are responsible for building up increasingly massive galaxies and for triggering starbursts and AGN activity \citep[e.g.,][]{toom77mergers,sand88ULIRGs,kauf93mergers,miho06mergers,hopk06merger,hopk08evolve,lotz11mergerrate,some15formationrev}. \citet{sand88ULIRGs} presented a basic framework in which the collision of two gas-rich disk galaxies would funnel gas towards the center of the system via tidal streams or shocks, thus creating a dusty, gas-rich environment to foster rapid star formation \citep[e.g.][]{lons06ulirg}. This dusty starburst stage would be selected as a ULIRG. As gas is fueling rapid star formation, it is continuously being funneled into the nucleus and also being accreted onto the black hole, thus also triggering AGN activity \citep[e.g.][]{hopk06merger,hopk08evolve}. Within this framework, gas from the galaxy can be expelled by a blowout phase driven by violent, dissipative feedback. 

\begin{figure}%[h]
    \centering
    \includegraphics[width=0.5 \textwidth]{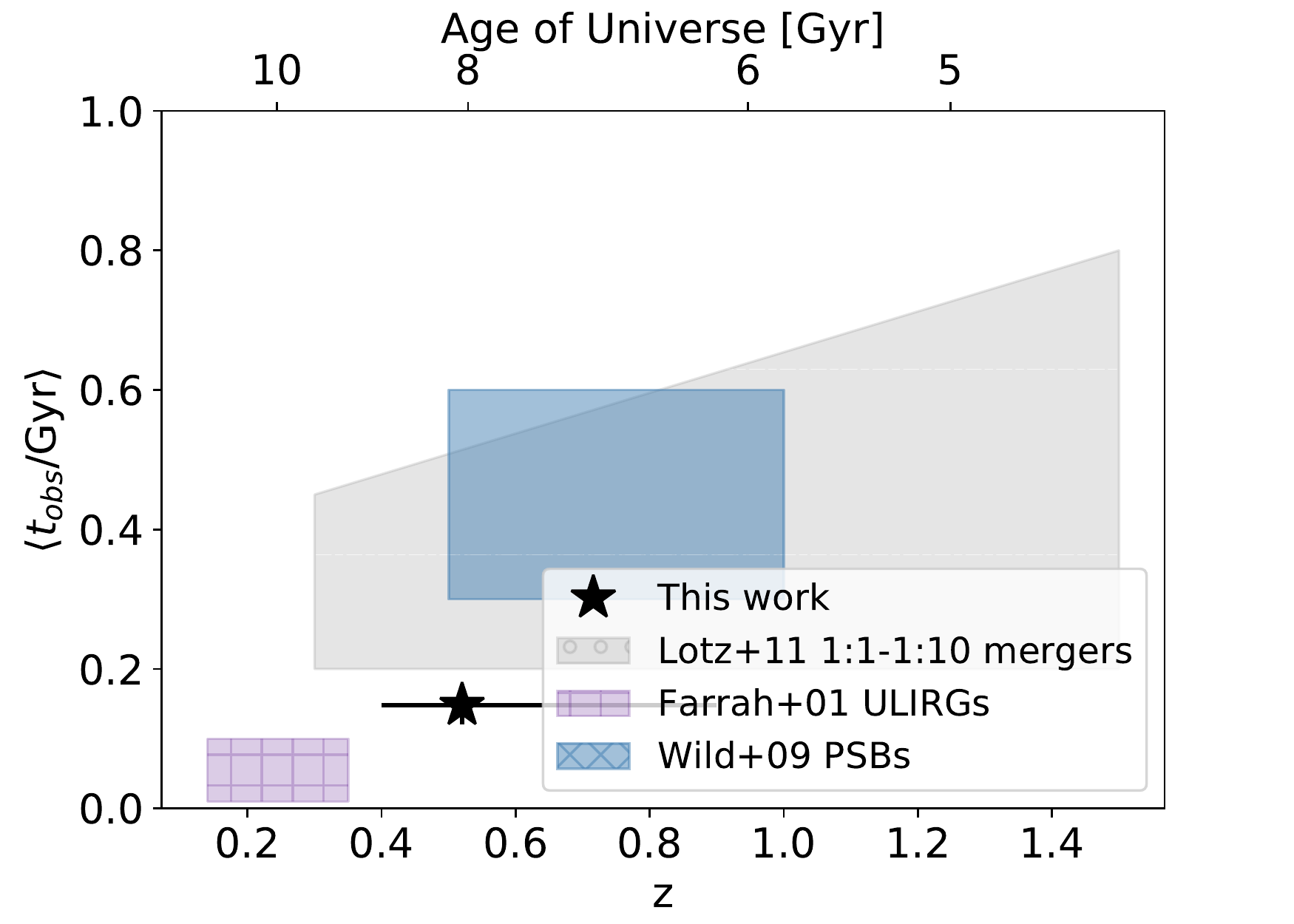}
    \caption{Comparison of the average timescales (in Gyr) upon which various phases of massive galaxy evolution would be observable. The black star represents the average selectability timescale for the modeled compact starburst galaxies in our sample, and its error bar along the redshift axis represents the size of the redshift range of our sources and the error bar along the $t_{obs}$ axis is the statistical uncertainty calculated via bootstrapping as described in Section \ref{sec:spacedensity} The grey, purple, and blue shaded regions represent the range of observable timescales for galaxy mergers \citep{lotz11mergerrate}, ULIRGs \citep{farr03ULIRGburstagn}, and post starburst galaxies (PSBs; \citealt{wild16PSBs}), respectively. We note that the timescales presented for galaxy mergers and PSBs correspond to the amount of time a source would be targeted under a set of selection criteria (similar to the value calculated for our sources), while the timescale for ULIRGs reflects the amount of physical time a source would experience star formation characteristic of the ULIRG phase. We elaborate on how we obtain the timescale estimates for the shaded regions in the text. It is clear that compact starburst galaxies like the ones in our sample occur on relatively short lived timescales that are comparable to that of ULIRG star formation.}
    \label{fig:timescales_compare}
\end{figure}

\begin{figure*}%[h]
    \centering
    \includegraphics[width=0.9\textwidth]{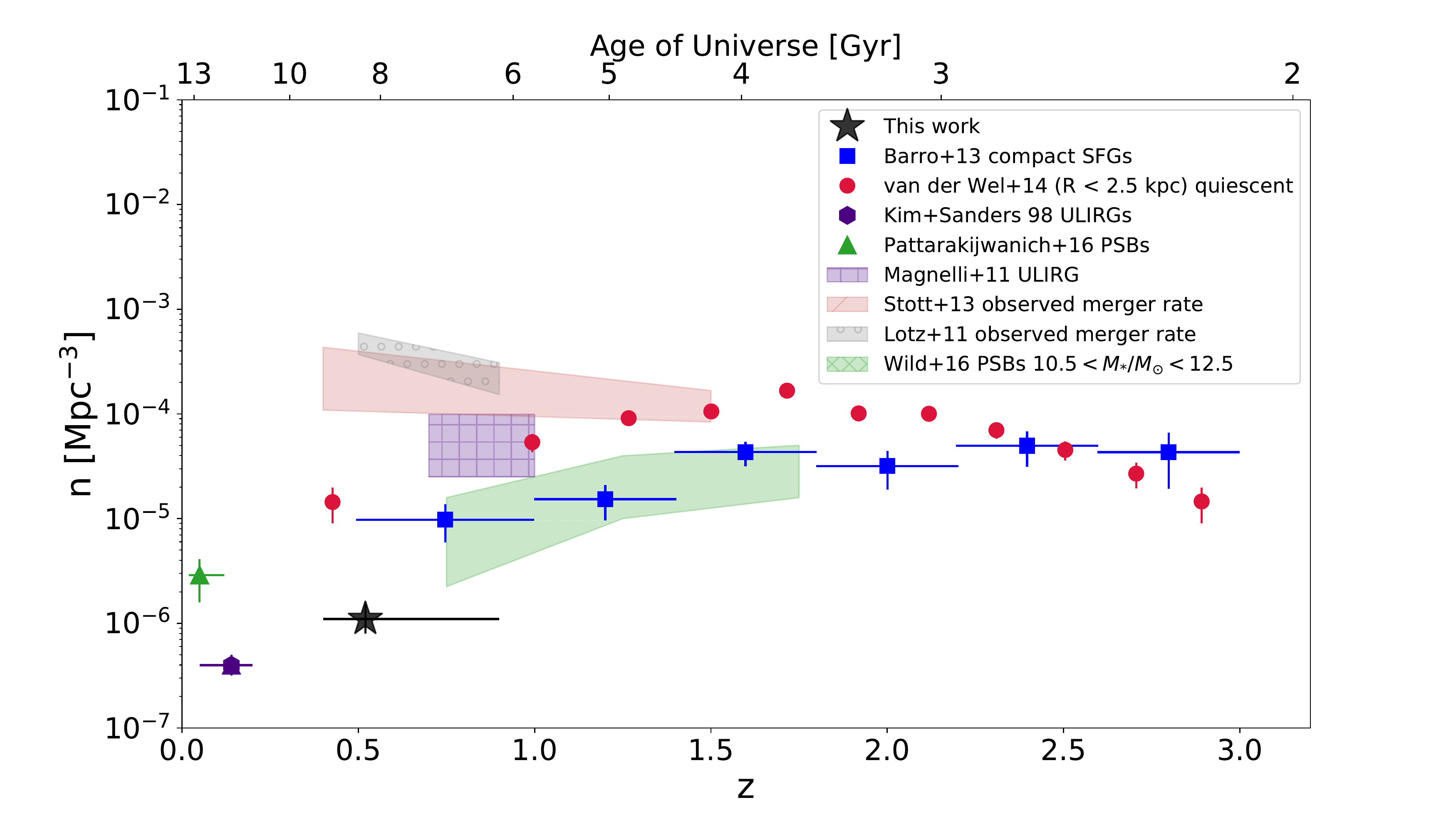}
    \caption{Comparison of the space densities of various phases of massive galaxy evolution. The black star represents the modeled space density for compact starburst galaxies like those in our observed sample. Its error bar along the redshift axis represents the size of the redshift range of our sources and the error bar along the space-density axis is the statistical uncertainty calculated via bootstrapping as described in Section \ref{sec:spacedensity}. We note that there are additional systematic errors, including uncertainty with model assumptions, which make this statistical error a lower limit. The blue squares represent the space density evolution of massive, compact star forming galaxies from the CANDELS survey \citep{barr13CANDELScompact}, the red points represent massive ($\log M_{*}/M_{\odot} \sim 11$), compact quiescent galaxies \citep{vand14highzcompact}, the green triangle represents low-$z$ PSBs \citep{patt16lowzPSBs}, and the purple hexagon represents low-$z$ ULIRGs \citep{kim98IRASulirgs}. The grey, red, purple, and green shaded regions depict the \citet{lotz11mergerrate} observed merger rate density, the \citet{stot13hizels_merger} observed merger rate density (calculated using merger observability timescales), ULIRG space density \citep{magn11IRlumfunc}, and intermediate-$z$ PSB space density \citep{wild16PSBs} ranges, respectively. The \citet{barr13CANDELScompact} points, \citet{lotz11mergerrate} region, and  \citet{stot13hizels_merger} region have been adjusted to account that our sources have masses $\log M_{*}/M_{\odot}> 10.5$, while most of the other populations shown include galaxies $\log M_{*}/M_{\odot}> 10$. While only a relatively small fraction of intermediate-$z$ major mergers will result in an extreme compact starburst similar to those in our sample, it is likely that sources like ours are the more extreme, lower-$z$ analogs to compact star forming galaxies more common in the early Universe and are closely related to intermediate-$z$ PSBs.}
    \label{fig:space_density_compare}
\end{figure*}

The galaxies in our observed sample have many features that could tie them into this evolutionary framework. We know that the galaxies for which we have \textit{HST} observations have disturbed morphological features such as tidal tails or two nuclei, which is indicative of them having undergone a recent merger \citep[e.g][]{sell14CSoutflow}. In addition to having disturbed morphologies, our galaxies host high velocity ionized and molecular gas outflows which can extend out to kpc scales \citep[e.g.][]{trem07CSoutflow,diam12CSstarform, geac13CSmolec,geac14CSstellarfeedback,sell14CSoutflow,geac18molecoutflow} or even over 100~kpc scales \citep{rupk19makani}. %It is difficult to conclusively characterize our sample of galaxies as ULIRGs as we do not have far-IR measurements, but it is entirely possible that our sample of galaxies and ULIRGs are related to one another and that our sample could consist of ULIRGs during their most extreme bursts of star formation and/or at the beginning of their blowout phases. 

In order to understand the evolutionary significance of extreme, compact star formation events like those observed in our galaxies, we need to contextualize their space density relative to that of various phases within massive galaxy, merger-driven evolution. Our results are summarized in Figures \ref{fig:timescales_compare} and \ref{fig:space_density_compare}, and we discuss in greater detail within this section.

%%% compact galaxies
\subsection{Evolution of massive compact galaxies}
The sample of galaxies we have been studying is comparable to a high-$z$ population of similarly compact, massive forming galaxies. Massive, quiescent galaxies in the Universe at $z>1.5$ are typically more compact than their local counterparts by roughly a factor of 5 \citep[e.g.][]{zirm05highzcompact, vand08highzcompact,vand14highzcompact}. The progenitors of these galaxies were likely compact star forming galaxies that were formed in gas-rich mergers of disk galaxies and were then rapidly quenched via some dissipative feedback, a formation scenario that is reminiscent of what we expect for ULIRGs and quiescent galaxies in the lower-$z$ Universe \citep[e.g.,][]{barr13CANDELScompact,stef13compactprogen, vand15massivecompact}. 

\citet{barr13CANDELScompact} observed populations of compact quiescent and star forming galaxies in the redshift range $\sim 1 < z <3$ to understand the evolutionary pathways that lead to the assembly of massive, compact quiescent galaxies we see predominantly in the early Universe. We include their compact star forming galaxy space density evolution as the blue squares in Figure \ref{fig:space_density_compare} for comparison with the intermediate-$z$ massive, compact starburst galaxies we are studying (black star). We adjust the points from \citet{barr13CANDELScompact} using redshift appropriate stellar mass functions \citep{mous13primus,adam21SMF} to account for the fact that their sample consists of sources with a wider stellar mass distribution than our sample. The adjusted space density is given as \begin{equation}
n_{\text{adjusted}} = n_{\text{literature}} \times \frac{\int ^{\infty}_{\text{lim, us}} \phi _{\text{SMF}} \ \text{d} \log M_{*}}{\int ^{\infty}_{\text{lim, lit}} \phi _{\text{SMF}} \ \text{d} \log M_{*}},\end{equation}
where $n_{\text{literature}}$ is the literature space density calculated for a larger mass range than our sample, and $\phi _{\text{SMF}}$ is the stellar mass function. We use the \citet{mous13primus} and \citet{adam21SMF} SMFs for $z \leq 1.5$ and $z > 1.5$, respectively. The \citet{barr13CANDELScompact} compact star forming galaxies have constant space densities at high redshift, but begin to decline at $z < \sim 1.5$. This decline is consistent with the decline in galaxy merger, star formation, and cold gas densities with decreasing redshift \citep[e.g.,][]{tacc10gasfrachiz,dadd10gasfrac,tacc13gasscaling,mada14cosmicSFH, riec19coldgas}. %Extrapolating this trend out to the $0.35 < z < 0.9$ range where our sources reside shows consistency with the space density we estimate for our extremely compact, massive starburst galaxies. 

%One of the most striking aspects of Figure \ref{fig:space_density_compare} is the consistency between the \citet{barr13CANDELScompact} space density evolution and the space density of our $z\sim 0.5$ compact starbursts. Our sources lie perfectly along that trend, suggesting that they could be the lower redshift analogs of the \citet{barr13CANDELScompact} compact star forming galaxies.  %%% want to discuss timescales but that will take some more time

We show in Figure \ref{fig:space_density_compare} that the space density of our sources lies only slightly below the space density evolution trend shown with the \citet{barr13CANDELScompact} compact star forming galaxies. We note that our galaxies are more extreme than the \citet{barr13CANDELScompact} sources as they are both more compact and more rapidly star forming. This likely biases our compact starburst space density to be slightly lower than that for the \citet{barr13CANDELScompact} galaxies. It is possible that our sources represent the low redshift analogs for an extreme subset of compact starburst galaxies that are more prevalent in the early Universe.

 Understanding how stellar feedback rapidly quenches star formation at intermediate redshift is necessary to be able to build models for galaxy formation and evolution in the early Universe when compact star formation events were significantly more common. For compact star-forming galaxies in the early Universe, it is difficult to observe the effects of feedback due to their high redshift and the fact they are commonly obscured by dust, making it nearly impossible to observe UV spectral signatures of outflows \citep[e.g.,][]{vand15massivecompact}. The broad consistency between the space density of our extreme, compact starburst galaxies and the \citet{barr13CANDELScompact} sample allows us to better understand how compact star formation might be a phase that massive galaxies go through across a wide range of cosmic time. 

\citet{barr13CANDELScompact} also presented a schematic representation of how galaxies evolve onto the local size-mass relation. Within this framework, compact star forming galaxies will experience rapid quenching via AGN or star formation feedback, resulting in a massive, compact quiescent galaxy population. Over cosmic time, these sources will undergo minor and major mergers resulting in a buildup of mass and size \citep[e.g.][]{naab09sizemass}. If our sources are the low-redshift analogs of early Universe compact star forming galaxies beginning their quenching phase, we would expect that they would also end up as compact, quiescent galaxies. We show the space density evolution from \citet{vand14highzcompact} for high-$z$, massive ($M_{*} \sim 10^{11} \ M_{\odot}$), compact ($R/(M_{*}/M^{11})^{0.75} < 2.5$ kpc) galaxies as red points in Figure \ref{fig:space_density_compare}. The space density of compact quiescent galaxies peaks just as that of compact star forming galaxies begins to decline. It then wanes with decreasing redshift due to size buildup via galaxy mergers. Within the lowest redshift bin, the \citet{vand14highzcompact} sources have a space density of $\sim 10$ larger than that of our compact, starburst galaxies. It is also worth noting that the compact quiescent galaxies would be considered to be ``compact'' for $\sim 2$ Gyr before minor mergers significantly contribute to size buildup \citep[e.g.,][]{naab09sizemass,newm12sizegrowth}--- a timescale that is significantly longer than the $\sim 100$ Myr timescale for which our sample would be targeted as extremely compact starbursts \citep[e.g.][]{barr13CANDELScompact}. In addition to this, the effective radii for the \citet{vand14highzcompact} sources is significantly larger than that of our nuclear starbursts. This could be due to the compact quiescent radii being more linked to the stellar mass profiles, while ours might be biased to small values because of mass-to-light ratio (M/L) effects. However, \citet{diam21CSoutflows} showed that even accounting for M/L effects that the stellar mass effective radius for our systems is on the order of 0.1-0.5 kpc, which indicates that our population could be even smaller and potentially more extreme than the compact quiescent galaxies in the \citet{vand14highzcompact} sample. All of this together suggests that a significant fraction of massive, compact quiescent sources at intermediate redshift could have recently gone through a starburst similar to what we observe for the galaxies in our sample.

%%% PSBs
\subsection{Comparison to post starburst galaxies}
In order to get a full picture of the role intermediate-$z$, extremely compact starbursts galaxies play in the buildup of a massive, quiescent population, we also need to understand the evolutionary stages that follow their bursts. By design of our selection criteria, the compact starburst galaxies in our sample are similar to PSBs in that they have B and A-star dominated spectral features and weak nebular emission. Understanding the population of PSBs in a similar redshift interval as our sources would provide context for quenching timescales as well as what the progenitors of PSBs might look like. 

\citet{wild16PSBs} studied a population of massive, PSBs within $0.5<z<2$, and determined that PSBs are a relatively short-lived, transitory phase in galaxy evolution, likely lasting $\sim 0.1 - 1$ Gyr \citep[see also][]{wild09PSBs}. This timescale range was determined by modeling PSBs in both toy-model and hydrodynamic simulations, and evolving them to determine the amount of time they would be targeted as PSBs--- a similar method to what we do here for our compact starburst galaxies. The PSBs selectability timescale is given as the blue region in Figure \ref{fig:timescales_compare}. Our compact starburst galaxies with selectability timescales of $\sim 100$ Myr would be selected for $10-100$\% of the time PSBs would be selected by their respective selection criteria.

It would be expected that extremely compact starburst galaxies and PSBs would have similar space densities within a given redshift range if they were two evolutionary stages that were directly related to each other. In other words, if compact starburst galaxies are the immediate progenitors to PSBs, they should be found in similar abundances. This is what is seen in Figure \ref{fig:space_density_compare}. The \citet{wild16PSBs} PSBs within the mass range $10.5 < \log M_{*}/M_{\odot} < 12.5$ show a decrease in space density with decreasing redshift. The lowest redshift bin for the \citet{wild16PSBs} PSBs overlaps with the upper limits of the redshift range probed for our compact starburst galaxies. The mass bin for \citet{wild16PSBs} is consistent with that of our sources so we did not have to correct for integrating the SMF within different mass intervals. Our sources overlap within the margin of error with the estimated PSB space density at the lowest redshift included in the \citet{wild16PSBs} sample.

The redshift evolution of the \citet{wild16PSBs} PSB space density is also consistent with declining star formation and cold-gas densities over cosmic time--- properties that would also impact the frequency of extremely compact bursts of star formation \citep[e.g][]{mada14cosmicSFH,riec19coldgas}. Since the cosmic SFR density sharply declines at low-$z$, we also want to compare our compact starburst space density to that of low-$z$ PSBs to determine if our calculated space density is consistent with the decline in PSB space density on the interval $0<z<1$. We calculate the $z\sim0.05$ PSB space density by integrating the lowest-$z$ luminosity function presented in \citet{patt16lowzPSBs}. This luminosity function is given per [5000 \AA] magnitude, a fiducial top hat filter used to calculate average $f_{\lambda}$ across $4950 < \lambda/\text{\AA} < 5100$ for the rest frame spectra of the PSBs in their sample. In order to calculate a comparable space density from this, we needed to construct a [5000 \AA] mass-luminosity relation to determine our bounds of integration. We did this by calculating [5000 \AA] magnitudes from SDSS spectra for the low-$z$ PSBs studied in \citet{fren18clockingPSBs} using the methodology described in \citet{patt16lowzPSBs} and using MPA-JHU stellar masses \citep{brin04stellarmass,trem04massmetal}. We then integrated the \citet{patt16lowzPSBs} luminosity function within $10.5 < \log M_{*}/\text{M}_{\odot} < 11.5$, which corresponds to $-23.3 < [5000 \text{\AA}] < -21.3$, to obtain a low-$z$ PSB space density of $\sim (2.9 ^{+1.2}_{-1.3})  \times 10^{-6} \ \text{Mpc}^{-3}$. This is given as the green triangle in Figure \ref{fig:space_density_compare}. This is of the same order of magnitude of that for our $z\sim0.5$ compact starburst galaxies, which supports that a fraction of the most extreme PSBs might have undergone an extremely compact starburst phase like that observed in our galaxies.

%%% ULIRGs
\subsection{Comparison to ULIRGs}
 Within the framework of merger-driven galaxy evolution, it is likely that extremely compact starburst events are most relevant in the remnants of major, gas-rich mergers. We also know that major, gas-rich mergers can trigger strong bursts of dusty star formation which would be observed as a ULIRG with $L_{FIR} > 10^{12} \ L_{\odot}$. It is possible that sources like the massive, extremely compact starburst galaxies in our sample could represent the transition between the dust-obscured ULIRG and the beginning of a galaxy-scale blowout. Here, we compare the selectability timescale and space density of our compact starbursts to that of ULIRGs in order to contextualize their importance in merger-driven galaxy evolution.
 
 The timescales upon which a galaxy will experience ULIRG-like star formation are poorly constrained. On the low end, SN-driven winds could cut the lifetime of a single starburst in a ULIRG to 1-10 Myr \citep[e.g.,][]{thor00starburst_ev}. However, studies of ULIRGs with a wide  variety of morphologies have allowed the ULIRG lifetime to be estimated to be in the 0.1-1 Gyr range \citep[e.g.][]{farr01ulirgs,murp01ULIRGages,farr03ULIRGburstagn}. It is possible that this wide range of estimated ULIRG lifetimes is due to the fact that it is likely that a ULIRG undergoes multiple large bursts of star formation, allowing it to be selected as such on discontinuous time intervals \citep[e.g.,][]{bekk01mergers,farr01ulirgs}. \citet{farr03ULIRGburstagn} analyzed a population of 41 local ULIRGs and found that most of their sources would have lifetimes $10 \lesssim \text{Myr} \lesssim 40$. From all of the values quoted above, we assume that the lifetime of a ULIRG is $\sim 1-100$ Myr, and show this range as the purple shaded region in Figure \ref{fig:timescales_compare}. However, it is important to make the distinction that these timescales are more strongly related to the physical timescales of dusty star formation than to observable lifetimes caused by respective selection criteria as discussed in other sections. The post-peak SF ages for the MgII galaxies in our sample calculated in \citet{davi_inprep_mgii_winds} are better comparisons to the ULIRG lifetimes due to the fact that they are tied more to the physical properties of the galaxies. As stated earlier, \citet{davi_inprep_mgii_winds} calculated the average post-peak SF age of $\sim 70$ Myr, which is largely consistent our estimate that they would be able to be targeted for $\sim 148 ^{+27}_{-24}$ Myr. These timescales are of a similar order of magnitude to that of ULIRGs, which is largely unsurprising because both types of systems are characterized by their energetic starbursts, albeit ours are a bit more extreme.
 
 We next compare our estimated compact starburst space density to that of ULIRGs in a similar redshift interval. \citet{kopr17irlumfunc} computed the evolution of the far-IR luminosity function for galaxies out to $z\sim5$. We estimate the observed space density of ULIRGs by adopting the $0.5 < z <1.5$ far-IR luminosity function presented here. Integrating the luminosity function for $L_{\text{IR}} > 10^{12} \ L_{\odot}$  gives $n_{\text{ULIRG}} \sim 6 \times 10^{-5} \ \text{Mpc}^{-3}$. This is shown as the purple shaded region in Figure \ref{fig:space_density_compare}, where the range of values is due to the uncertainty in the Schechter function fit as described in \citet{kopr17irlumfunc}. We note that we do not correct for differences in the mass distributions between the ULIRG sample and our sources because ULIRG sample was luminosity selected. Similarly, \citet{magn09farIR} calculated the evolving far-IR luminosity function and space density for ULIRGs for several redshift bins within the interval $0.4<z<1.3$. For the $0.4<z<0.7$ and $0.7<z<1$ bins, $n_{\text{ULIRG}} \sim 3 \times 10^{-5} \ \text{Mpc}^{-3}$ and $n_{\text{ULIRG}} \sim 2 \times 10^{-5} \ \text{Mpc}^{-3}$, respectively. 
 
 Comparing these values to our estimated compact starburst space density ($(1.1 ^{+0.5}_{-0.3})  \times 10^{-6} \ \text{Mpc}^{-3}$) suggests that it is possible that $\sim 3-8$\% of intermediate-$z$ ULIRGs can experience a phase similar to that observed in our sample of extremely compact starburst galaxies. The physical timescales of ULIRGs and our compact starbursts are driven by the same processes, and they are on the same order of magnitude, while there is a factor $\sim12-40$ difference in their space densities. It is possible the sources in our sample represent a small fraction of the most extreme population of ULIRGs that have the highest SFRs and/or are the most compact. 
 
We also compare the space density of our intermediate-$z$ massive, compact starburst galaxies to that of low-$z$ ULIRGs, similar what we hae done in the previous subsection for PSBs since we expect a sharp decline in the ULIRG space density alongside that of the cosmic SFR density \citep[e.g.,][]{mada14cosmicSFH}. \citet{kim98IRASulirgs} presented a luminosity function for $0.05<z<0.2$ ULIRGs, and integrating the luminosity for $\log \ L_{\text{IR}}/\text{L}_{\odot} > 12$ gives a space density of $\sim (4 \pm 1) \times 10^{-7} \ \text{Mpc}^{-3}$. This is given as the purple hexagon in Figure \ref{fig:space_density_compare}. Given that the space density of our intermediate-$z$, compact starburst galaxies is calculated in a redshift range between that of the low and intermediate-$z$ ULIRGs, this very steep decline in ULIRG space density also suggests that a small fraction of ULIRGs could undergo a phase like that observed in our galaxies as they evolve.

 \subsection{Comparison to $z\sim 0.5$ merger rate per co-moving unit volume}
 Since extremely compact starburst galaxies are likely formed by the merging of gas-rich disk galaxies, it is important to characterize how many major mergers could produce events like those observed in our sample of galaxies. This requires having knowledge of the major merger rate over a given redshift range. In the past few decades, much work has been done to constrain the galaxy-galaxy merger rate throughout cosmic time. However, there are large systematic uncertainties in this measurement that have prevented the reaching of a consensus between theory and observations and even between different observational techniques. Here, we summarize the most recent results in calculating the $z\sim 0.5$ galaxy merger rate per co-moving unit volume and use them to contextualize our compact starburst space density. To be more concise, we will refer to the merger rate per co-moving unit volume as the merger rate density for the rest of this paper.
 
A crucial piece of calculating the galaxy merger rate density is understanding the timescales upon which a system would be identified as a major merger. This is also the aspect of the calculation that contributes the most uncertainty to the major merger rate density. The two main methods to identify merging galaxies are to select systems with disturbed morphologies \citep[e.g.,][]{abra94morphclass,abra03SDSSmorph,cons09mergers,lotz08timescales} or to search for systems comprised of close pairs \citep[e.g,][]{lefe00HSTmergers,bluc09closepairs}. Each of these methods probe different stages of the merger and are susceptible to different biases. Close pair selection identifies sources before the merger begins but morphological selection can detect systems before, during, and after the merger occurs, allowing morphologically selected galaxy mergers to be identifiable on different timescales than their close pair counterparts. 

In Figure \ref{fig:timescales_compare}, we compare the selectability timescale calculated for our modeled compact starburst galaxies (black star) to that of all galaxy mergers presented in \cite{lotz08timescales} (grey shaded region). The \citet{lotz11mergerrate} region reflects the range of timescales calculated for simulated systems with mass ratios $1:10 < \mu < 1:1$ that were selected morphologically (for a detailed review; \citealt{abra94morphclass,abra03SDSSmorph,lotz11mergerrate}). We find that extreme compact starburst events are selectable for a fraction of the amount of time that a morphologically selected galaxy merger would be under its own respective criteria. 

%However, we note that the timescales presented in \citet{lotz11mergerrate} account for both major and minor mergers and that the timescales for disturbed morphologies due to only major mergers would be even longer than shown here.

Having constraints on galaxy merger timescales allows for the merger rate density to be calculated. We show our calculated compact starburst space density (black star) in conjunction with merger rate densities (grey and red shaded regions) as well as the space densities of other phases of merger-driven evolution in Figure \ref{fig:space_density_compare}. The grey shaded region represents the range of the predicted observable merger rate densities calculated in \citet{lotz11mergerrate}, and the red shaded region represents the observed range of merger rate densities presented in \citet{stot13hizels_merger} which used \citet{lotz11mergerrate} predicted observable timescales. Both the \citet{lotz11mergerrate} and \citet{stot13hizels_merger} merger rate densities were calculated for samples containing galaxies with $\log M_{*}/M_{\odot} > 10$, while the compact starburst galaxies in our sample are typically  $\log M_{*}/M_{\odot} > 10.5$. We therefore adjusted the \citet{lotz11mergerrate} and \citet{stot13hizels_merger} merger rate densities to ensure that we are working within the same mass interval of the galaxy stellar mass function (SMF) within the appropriate redshift range, as described above. We also converted these merger rate densities to merger space densities by assuming a typical merger timescale of 0.5 Gyr \citep{lotz11mergerrate}.

We find that our estimated massive compact starburst space density is $\sim 200$ times smaller than the merger rate density within a similar redshift interval, suggesting that only a small fraction of galaxy mergers would trigger an extreme burst of compact star formation similar to our observed sample. However, we reiterate that the \citet{lotz11mergerrate} and \citet{stot13hizels_merger} merger rates consider both major and minor mergers. It is likely that these compact starburst events are triggered only by gas-rich major (mass ratio 1:1 - 4:1)
 mergers which only make up a fraction of the total number of mergers occurring across a given redshift range \citep[e.g.,][]{lin10_mergerenv}. This suggests that although only a small fraction of all galaxy mergers might result in extremely compact starbursts, that these could be a likely result of a larger fraction of gas-rich major mergers.

 \subsection{Comparison to $z\sim 0.5$ massive, quiescent galaxies}
 Another way of understanding the role of compact starburst galaxies in the buildup of quiescent galaxy populations is to compare their space density to that of massive, quiescent galaxies within the same redshift range. \citet{mous13primus} presented a detailed study of galaxies targeted in PRism Multi-object Survey (PRIMUS) and provided contsraints on the evolution of the stellar mass function from $0 < z < 1$. The galaxies in PRIMUS were sorted into star forming and quiescent populations, and the evolution of their space density was calculated across different stellar mass and redshift bins. For quiescent PRIMUS galaxies in the mass range $10.5 < \log M/M_{\odot} < 11$, their space density increases by $\sim2 \times 10^{-4} \ \text{Mpc}^{-3}$ from $z \sim 0.8$ to $z \sim 0.35$. The net decline in space density for star forming galaxies in this redshift interval is $\sim 9\times 10^{-5} \ \text{Mpc}^{-3}$. These changes in space density are comparable to the merger rate in this redshift range and are a factor of $\sim 1000$ larger than our measured space density of $n\sim (1.1 ^{+0.5}_{-0.3}) \times 10^{-6} \ \text{Mpc}^{-3}$ for our sample of massive, compact starburst galaxies. This is broadly consistent with short-lived compact starbursts existing for $\sim 100$ Myr, evolving into massive, quiescent galaxies which would exist on $\sim$Gyr timescales. It is likely that this is a relatively rare phase of galaxy evolution within the general population of massive, quiescent galaxies. However, it is possible that the fraction of those that have also previously undergone extreme ULIRG or PSB phases also could have experienced extremely, compact starbursts like those in our sample.

\section{Summary \& Conclusions}

In order to build up a population of quiescent galaxies, otherwise gas-rich and star forming galaxies need to undergo some type of quenching process to either disrupt or expel the gas in the system. Violent, dissipative feedback in which either AGN activity or rapid star formation injects energy into the ISM is an important process that impedes the formation of stars in a galaxy. Observationally, feedback manifests as large-scale gas outflows being driven from a galaxy. %Although we know that feedback is imperative to galaxy quenching, the importance of AGN activity versus star formation is still up for debate.

Within the context of merger-driven galaxy evolution, we expect gas-rich mergers of massive star forming galaxies to trigger dusty starburst events that would then be followed by a blowout event in which nuclear gas and dust is expelled from the system, therefore exposing the nuclear regions of the galaxy. In this work, we have studied a population of 115 $z\sim0.5$ massive galaxies that are experiencing extreme, compact starburst events and outflows. Resolved \textit{HST} WFC3 observations of a subset of these show that they are merger remnants, suggesting that these types of events could be an phase within a simple merger-driven evolutionary pathway. 

Our goal for this work was to determine how long galaxies like the ones we observe would be selected under a certain set of selection criteria, to estimate their space density, and to place them into cosmological context with other evolutionary phases massive galaxies could experience. We do this by empirically modeling the stellar populations of $z\sim0.5$ massive, compact starburst galaxies. Our model is dependent on four parameters: nuclear burst age, burst mass fraction, optical depth of dust enshrouding newly formed stars, and total galaxy stellar mass. These posterior distributions for these parameter values are constrained for each of the 115 galaxies in our sample by fitting the SDSS $ugriz$ and \textit{WISE} W1/W2 photometry for the 151 galaxies in our sample using an MCMC technique. We randomly draw sets of parameters from the Gaussian models for the MCMC-calculated posterior distributions to assemble a mock population of compact starburst galaxies. We evolve the modeled sources to determine the timescales under which the galaxies we model would be selected by our targeting criteria. We find that this timescale is $148 ^{+27}_{-24}$ Myr and that the corresponding intrinsic space density is $n_{\text{CS}}\sim (1.1 ^{+0.5}_{-0.3}) \times 10^{-6} \ \text{Mpc}^{-3}$.

%We constructed an empirical model that allowed us to assemble a large population of r $z\sim0.5$ compact starburst galaxies and evolve them to determine a timescale under which they would be selected by our criteria. We calculated the average selectability timescale of our galaxies to be $\sim 148 ^{+27}_{-24}$ Myr and the space density to be $n\sim (1.1 ^{+0.5}_{-0.3}) \times 10^{-6} \ \text{Mpc}^{-3}$.

Our results, as summarized in Figure \ref{fig:space_density_compare}, suggest that our observed population of extreme compact starburst galaxies could fit into an evolutionary scheme described in \citet{barr13CANDELScompact}. At higher redshifts massive, compact star forming galaxies are more common, and they are believed to be the progenitors of massive, compact quiescent galaxies. Based on comparisons with the \citet{barr13CANDELScompact} sample of massive, compact galaxies it is likely that our sources follow a similar life cycle in which a gas-rich major merger triggers a burst of star formation. This starburst then drives massive, high velocity gas outflows, thus rapidly quenching the galaxy. This galaxy would be observable for $\sim 100$ Myr timescales as a PSB \citep[e.g.,][]{wild16PSBs}, and would then evolve into a massive, compact, quiescent galaxy. Throughout cosmic time, the massive, quiescent galaxy will undergo minor mergers, allowing it to grow in both mass and size to become a typical quiescent galaxy consistent with the mass-size relation of the massive quiescent galaxy population at z=0, which is notably devoid of compact quiescent galaxies \citep[e.g.,][]{tayl10compactquies}. Although it is more common for galaxies to experience this timeline earlier in the Universe, our galaxies appear to be consistent with these trends within their respective redshift interval. The space density of our massive, compact starbursts suggests that they can contribute to the buildup of a fraction of PSBs and massive, extreme compact quiescent galaxies within their epoch, which in turn could contribute to the overall population of massive, quiescent galaxies in the future.

\begin{acknowledgments}
We acknowledge support from the National Science Foundation (NSF) under a collaborative grant (AST-1814233, 1813299, 1813365, 1814159 and 1813702) and from the Heising-Simons Foundation grant 2019-1659. 
\end{acknowledgments}

%\bibliography{whalen_research}
 \newcommand{\noop}[1]{}

\appendix
\section{Auxillary MCMC Output}
%\restartappendixnumbering

\centering

\begin{longtable}{ccccccccccc}
\caption{Properties for the galaxies included in our sample.}
\label{tab:sample}\\
 SDSS ID & $z$ & $\langle \log M_{*}/M_{\odot} \rangle$ & $\sigma _{\log M_{*}/M_{\odot}}$ & SDSS $u$ & SDSS $g$ & SDSS $r$ & SDSS $i$ & SDSS $z$ & \textit{WISE} W1 & \textit{WISE} W2\\
   & & & & (AB)& (AB) & (AB) & (AB) & (AB) & (Vega) & (Vega)\\
 (1) & (2) & (3) & (4) & (5) & (6) & (7) & (8) & (9) & (10) & (11) \\
 \hline
 J1015+0004 & 0.417 & 11.0 & 0.07 & 22.03 & 20.71 & 19.25 & 18.95 & 18.77 & 15.83 & 15.38 \\
 J1109-0040 & 0.593 & 11.4 & 0.47 & 22.07 & 20.88 & 19.46 & 18.8  & 18.61 & 15.26 & 15.22 \\
 J1210+0030 & 0.441 & 11.1 & 0.08 & 21.88 & 20.87 & 19.37 & 19.02 & 18.79 & 15.87 & 15.3  \\
 J1341-0009 & 0.446 & 11.0 & 0.19 & 22.34 & 20.96 & 19.38 & 19.05 & 18.79 & 15.74 & 15.74 \\
 J1434-0052 & 0.461 & 11.3 & 0.51 & 23.45 & 21.04 & 19.29 & 18.66 & 18.31 & 14.86 & 14.64 \\
 J1440+0039 & 0.564 & 11.2 & 0.10 & 20.93 & 20.4  & 19.27 & 18.86 & 18.74 & 15.59 & 15.59 \\
 J1125-0145 & 0.519 & 10.9 & 0.27 & 19.6  & 19.33 & 18.69 & 18.48 & 18.39 & 14.84 & 14.65 \\
 J0745+3754 & 0.406 & 10.7 & 0.22  & 20.27 & 19.86 & 19.14 & 18.79 & 18.46 & 14.78 & 14.13 \\
 J0251-0657 & 0.406 & 11.1 & 0.27 & 22.91 & 21.14 & 19.39 & 18.88 & 18.57 & 15.38 & 15.19 \\
 J0905+5759 & 0.711 & 10.8 & 0.28 & 19.91 & 19.58 & 19.4  & 19.1  & 19.14 & 15.56 & 15.46 \\
 J1219+0336 & 0.451 & 11.0 & 0.21 & 20.15 & 19.52 & 18.79 & 18.53 & 18.33 & 14.99 & 14.56 \\
 J1232+0226 & 0.418 & 11.1 & 0.22 & 21.55 & 20.36 & 18.81 & 18.53 & 18.4  & 15.41 & 15.25 \\
 J1440+0107 & 0.456 & 10.9 & 0.23 & 20.63 & 20.26 & 19.38 & 18.97 & 18.76 & 15    & 14.53 \\
 J1441+0116 & 0.537 & 11.0 & 0.22 & 20.35 & 19.76 & 19.34 & 18.97 & 18.68 & 15.33 & 15.1  \\
 J0901+0314 & 0.459 & 10.6 & 0.23 & 19.55 & 19.29 & 18.82 & 18.7  & 18.57 & 15.22 & 15.01 \\
 J1107+0417 & 0.467 & 10.6 & 0.22 & 19.96 & 19.52 & 19.07 & 18.89 & 18.7  & 15.58 & 14.93 \\
 J1453+6022 & 0.406 & 10.9 & 0.15 & 20.49 & 20.04 & 19.02 & 18.78 & 18.55 & 15.61 & 15.33 \\
 J1506+6131 & 0.437 & 10.3 & 0.17 & 19.69 & 19.58 & 19.12 & 19.04 & 19.16 & 15.72 & 15.52 \\
 J1610+5104 & 0.469 & 11.1 & 0.07 & 22.1  & 20.93 & 19.35 & 18.92 & 18.76 & 15.68 & 15.51 \\
 J1635+4709 & 0.699 & 11.6 & 0.13 & 20.65 & 20.28 & 19.51 & 18.75 & 18.56 & 15.21 & 15.11 \\
 J2116-0634 & 0.728 & 11.3 & 0.18 & 20.74 & 20.02 & 19.72 & 19.2  & 19.05 & 15.51 & 15.55 \\
 J2311-0839 & 0.725 & 11.7 & 0.14 & 21.15 & 20.89 & 19.93 & 18.92 & 18.71 & 15.4  & 15.29 \\
 J2140+1209 & 0.751 & 11.1 & 0.25 & 20.63 & 20.19 & 19.85 & 19.31 & 19.1  & 15.57 & 14.98 \\
 J2256+1504 & 0.727 & 11.4 & 0.22 & 20.76 & 20.1  & 19.59 & 18.91 & 18.74 & 15.12 & 15.19 \\
 J2319+1435 & 0.422 & 10.5 & 0.36 & 22.62 & 21.07 & 19.42 & 19.01 & 18.78 & 15.77 & 15.44 \\
 J0826+4305 & 0.603 & 10.7 & 0.27 & 19.64 & 19.43 & 19.14 & 18.88 & 18.85 & 15.42 & 15.13 \\
 J0951+5514 & 0.402 & 11.3 & 0.11 & 20.65 & 20.01 & 18.91 & 18.51 & 18.15 & 14.85 & 14.37 \\
 J1235+6140 & 0.599 & 11.3 & 0.48 & 20.91 & 20.31 & 19.19 & 18.61 & 18.51 & 15.4  & 15.13 \\
 J1253+6256 & 0.536 & 10.4 & 0.17 & 19.69 & 19.64 & 19.3  & 19.25 & 19.22 & 16.16 & 15.68 \\
 J1506+5402 & 0.608 & 10.7 & 0.27 & 19.28 & 19.13 & 18.88 & 18.65 & 18.61 & 15.26 & 14.78 \\
 J1248+0601 & 0.632 & 11.2 & 0.18 & 20.89 & 20.33 & 19.49 & 18.98 & 18.85 & 15.77 & 15.64 \\
 J1117+5123 & 0.49  & 11.3 & 0.11 & 21.06 & 20.42 & 19.24 & 18.91 & 18.68 & 15.33 & 15.38 \\
 J1020+5331 & 0.457 & 11.0 & 0.31 & 22.53 & 20.68 & 19.21 & 18.88 & 18.69 & 15.77 & 15.62 \\
 J1401-0223 & 0.402 & 11.0 & 0.20 & 20.36 & 19.91 & 19.05 & 18.64 & 18.29 & 15.01 & 14.54 \\
 J0933+4135 & 0.441 & 10.7 & 0.25 & 19.07 & 18.97 & 18.46 & 18.39 & 18.24 & 15.14 & 14.59 \\
 J0939+4251 & 0.411 & 10.9 & 0.17 & 20.05 & 19.58 & 18.73 & 18.52 & 18.24 & 15.18 & 14.89 \\
 J1142+6037 & 0.568 & 11.5 & 0.29  & 20.86 & 20.13 & 18.81 & 18.29 & 18.17 & 15.05 & 14.79 \\
 J1713+2817 & 0.577 & 11.3 & 0.16 & 20.82 & 20.3  & 19.33 & 18.91 & 18.86 & 15.52 & 15.23 \\
 J1720+3017 & 0.684 & 11.6 & 0.10 & 21.25 & 20.67 & 19.75 & 18.89 & 18.78 & 15.45 & 15.47 \\
 J2118+0017 & 0.459 & 10.8 & 0.25 & 20.17 & 19.78 & 18.96 & 18.74 & 18.53 & 14.96 & 14.25 \\
 J0922+0452 & 0.476 & 11.1 & 0.08  & 21.22 & 20.34 & 18.99 & 18.79 & 18.57 & 15.7  & 15.49 \\
 J1052+0607 & 0.555 & 10.9 & 0.14 & 20.53 & 20.14 & 19.32 & 19    & 18.86 & 15.69 & 15.82 \\
 J1353+5300 & 0.408 & 11.3 & 0.18 & 20.43 & 19.84 & 18.81 & 18.38 & 18.12 & 14.65 & 14.2  \\
 J1436+5017 & 0.454 & 11.0 & 0.19 & 20.22 & 19.81 & 18.83 & 18.61 & 18.42 & 15.36 & 14.91 \\
 J1558+3957 & 0.402 & 10.6 & 0.23  & 19.37 & 19.07 & 18.54 & 18.44 & 18.24 & 15.17 & 14.55 \\
 J1604+3939 & 0.564 & 11.7 & 0.29 & 20.85 & 20.01 & 18.8  & 18.21 & 18.06 & 14.58 & 14.48 \\
 J0828+0336 & 0.572 & 11.0 & 0.11 & 20.9  & 20.3  & 19.3  & 18.98 & 18.94 & 16.09 & 15.77 \\
 J0808+2709 & 0.563 & 11.1 & 0.19 & 20.63 & 20.09 & 19.4  & 18.9  & 18.77 & 15.52 & 14.94 \\
 J1009+4336 & 0.519 & 10.9 & 0.26  & 19.6  & 19.37 & 18.79 & 18.56 & 18.38 & 14.98 & 14.62 \\
 J1133+0956 & 0.483 & 11.0 & 0.19  & 20.45 & 19.93 & 19.05 & 18.8  & 18.75 & 15.2  & 14.95 \\
 J0900+3212 & 0.496 & 11.3 & 0.22 & 20.18 & 19.84 & 18.96 & 18.5  & 18.14 & 14.67 & 14.49 \\
 J1330+4821 & 0.444 & 11.5 & 0.16  & 20.6  & 19.73 & 18.76 & 18.32 & 17.97 & 14.64 & 14.13 \\
 J1420+5313 & 0.742 & 11.8 & 0.20 & 20.72 & 20.39 & 19.84 & 19.01 & 18.67 & 14.68 & 14.61 \\
 J1556+4234 & 0.401 & 11.4 & 0.11 & 20.42 & 19.76 & 18.52 & 18.19 & 17.88 & 14.8  & 14.38 \\
 J1456+3849 & 0.421 & 10.8 & 0.23 & 19.84 & 19.48 & 18.88 & 18.49 & 18.16 & 14.44 & 13.8  \\
 J1459+3844 & 0.433 & 10.5 & 0.22 & 19.93 & 19.64 & 19.05 & 18.9  & 18.68 & 15    & 14.56 \\
 J1037+4048 & 0.439 & 11.1 & 0.16 & 22.17 & 20.94 & 19.4  & 18.98 & 18.74 & 15.74 & 15.56 \\
 J1248+4444 & 0.43  & 10.7 & 0.22 & 19.71 & 19.49 & 18.83 & 18.62 & 18.48 & 15.14 & 14.76 \\
 J1447+3650 & 0.414 & 11.0 & 0.06 & 22.71 & 20.89 & 19.17 & 18.84 & 18.72 & 15.7  & 15.58 \\
 J1520+3334 & 0.516 & 11.3 & 0.12 & 21.91 & 20.78 & 19.39 & 18.98 & 18.88 & 15.41 & 15.18 \\
 J1611+2650 & 0.483 & 11.4 & 0.13 & 20.97 & 20.18 & 18.97 & 18.67 & 18.45 & 14.92 & 14.62 \\
 J1039+4537 & 0.634 & 11.2 & 0.26 & 20.37 & 20    & 19.42 & 18.98 & 18.86 & 15.02 & 14.87 \\
 J1035+3854 & 0.422 & 11.0 & 0.29 & 22.44 & 20.8  & 19.26 & 18.96 & 18.75 & 15.85 & 15.59 \\
 J1052+4104 & 0.576 & 10.9 & 0.18 & 20.14 & 19.84 & 19.27 & 18.96 & 18.89 & 15.78 & 15.74 \\
 J1215+4233 & 0.479 & 11.2 & 0.33 & 22.22 & 20.86 & 19.21 & 18.82 & 18.58 & 15.48 & 15.22 \\
 J1244+4140 & 0.459 & 10.8 & 0.18 & 19.91 & 19.54 & 18.79 & 18.64 & 18.45 & 15.71 & 15.15 \\
 J0921+3251 & 0.73  & 11.1 & 0.41  & 26.53 & 15.87 & 14.65 & 16.92 & 16.62 & 15.01 & 14.96 \\
 J1012+1134 & 0.411 & 11.0 & 0.42 & 24.01 & 21.22 & 19.6  & 19.07 & 18.79 & 15.43 & 14.81 \\
 J1113+1119 & 0.628 & 11.6 & 0.63 & 20.57 & 17.48 & 17.04 & 16.68 & 16.5  & 15.49 & 15.43 \\
 J1232+0723 & 0.401 & 10.7 & 0.22 & 19.86 & 19.41 & 18.73 & 18.6  & 18.42 & 14.96 & 14.7  \\
 J1239+0731 & 0.542 & 11.0 & 0.18 & 20.51 & 20.12 & 19.29 & 18.95 & 18.85 & 15.62 & 15.49 \\
 J1415+4830 & 0.496 & 11.0 & 0.21 & 19.66 & 19.2  & 18.73 & 18.34 & 18.08 & 14.12 & 13.4  \\
 J1450+4621 & 0.782 & 11.6 & 0.15 & 20.6  & 20.09 & 19.66 & 18.89 & 18.85 & 15.24 & 15.23 \\
 J1658+2354 & 0.498 & 11.4 & 0.17  & 19.74 & 19.22 & 18.33 & 18.07 & 17.94 & 14.54 & 14.36 \\
 J0908+1039 & 0.502 & 11.0 & 0.23 & 19.77 & 19.45 & 18.74 & 18.47 & 18.27 & 14.89 & 14.6  \\
 J1119+1526 & 0.491 & 11.1 & 0.07 & 22.09 & 20.95 & 19.43 & 19.04 & 18.87 & 15.82 & 15.66 \\
 J0830+5552 & 0.526 & 11.0 & 0.25 & 20.19 & 19.85 & 19.16 & 18.81 & 18.56 & 14.82 & 14.49 \\
 J1435+0846 & 0.404 & 11.3 & 0.15 & 20.28 & 19.73 & 18.61 & 18.32 & 18.05 & 14.99 & 14.56 \\
 J0742+4844 & 0.431 & 11.0 & 0.14 & 20.79 & 20.19 & 19.03 & 18.83 & 18.59 & 15.6  & 15.19 \\
 J0752+1806 & 0.619 & 10.5 & 0.13 & 20.44 & 19.91 & 19.5  & 19.03 & 20.65 & 14.66 & 14.32 \\
 J0836+2526 & 0.531 & 10.8 & 0.23 & 20.75 & 20.29 & 19.28 & 18.94 & 18.8  & 16.12 & 15.82 \\
 J1016+3026 & 0.402 & 10.8 & 0.30 & 23    & 20.8  & 19.18 & 18.83 & 18.64 & 15.72 & 16.02 \\
 J1133+3958 & 0.487 & 11.1 & 0.21 & 19.7  & 19.29 & 18.52 & 18.29 & 18.15 & 14.74 & 14.46 \\
 J1229+3545 & 0.614 & 11.4 & 0.41 & 20.57 & 20    & 19    & 18.46 & 18.33 & 15.18 & 15.1  \\
 J1301+3615 & 0.573 & 11.3 & 0.19 & 20.51 & 20.01 & 19.13 & 18.68 & 18.59 & 15.05 & 14.91 \\
 J0901+2338 & 0.438 & 10.8 & 0.23 & 20.06 & 19.45 & 18.96 & 18.58 & 18.39 & 14.39 & 13.7  \\
 J0911+2619 & 0.471 & 11.0 & 0.24 & 20.18 & 19.71 & 18.95 & 18.58 & 18.36 & 14.5  & 13.92 \\
 J1403+2440 & 0.455 & 11.0 & 0.15 & 22.2  & 20.79 & 19.24 & 18.93 & 18.76 & 15.77 & 15.69 \\
 J1505+2312 & 0.417 & 11.0 & 0.07 & 22.6  & 20.89 & 19.33 & 18.98 & 18.71 & 15.78 & 15.39 \\
 J1548+1834 & 0.688 & 11.2 & 0.22  & 20.53 & 20.08 & 19.55 & 18.93 & 18.89 & 15.37 & 15.31 \\
 J1634+1729 & 0.491 & 10.9  & 0.23  & 20.71 & 20.1  & 19.35 & 19.04 & 18.78 & 15.22 & 14.54 \\
 J1635+1749 & 0.469 & 11.0 & 0.24  & 20.71 & 20.17 & 19.21 & 18.93 & 18.75 & 15.08 & 14.67 \\
 J1226+2753 & 0.427 & 10.3 & 0.14 & 19.14 & 19.14 & 18.83 & 18.81 & 18.78 & 15.87 & 15.46 \\
 J0936+2237 & 0.571 & 11.3  & 0.49 & 22.66 & 21.12 & 19.48 & 18.86 & 18.63 & 15.41 & 15.34 \\
 J1012+2258 & 0.504 & 11.5 & 0.20 & 20.74 & 20.13 & 18.81 & 18.43 & 18.21 & 14.89 & 14.69 \\
 J1000+2816 & 0.469 & 11.2 & 0.14  & 20.74 & 20.25 & 19.17 & 18.76 & 18.56 & 15.22 & 15.22 \\
 J0941+1827 & 0.569 & 11.5 & 0.14  & 21.02 & 20.26 & 19.1  & 18.61 & 18.36 & 15.2  & 14.97 \\
 J1005+1836 & 0.402 & 10.8 & 0.33 & 24.96 & 21.13 & 19.48 & 19.02 & 18.79 & 15.7  & 15.56 \\
 J0912+1523 & 0.747 & 11.7 & 0.29 & 20.91 & 20.37 & 19.59 & 18.64 & 18.4  & 15.23 & 15.06 \\
 J0900+1130 & 0.407 & 11.2 & 0.21 & 20.64 & 19.97 & 19.04 & 18.62 & 18.18 & 14.66 & 14.04 \\
 J1203+1807 & 0.595 & 11.4 & 0.38 & 22.37 & 21.25 & 19.73 & 18.96 & 18.82 & 15.41 & 15.38 \\
 J1205+1818 & 0.526 & 10.6  & 0.27 & 19.01 & 18.88 & 18.54 & 18.41 & 18.45 & 15.19 & 14.84 \\
 J1256+1826 & 0.424 & 11.0 & 0.39 & 22.52 & 21.02 & 19.35 & 18.88 & 18.6  & 15.42 & 15.27 \\
 J1248+1954 & 0.561 & 11.0 & 0.17 & 20.15 & 19.8  & 19.13 & 18.81 & 18.79 & 15.68 & 15.65 \\
 J1352+1653 & 0.533 & 11.3 & 0.32 & 22.07 & 21.07 & 19.47 & 18.88 & 18.63 & 15.57 & 15.36 \\
 J1400+1524 & 0.564 & 11.3 & 0.15  & 20.72 & 20.35 & 19.35 & 18.86 & 18.76 & 15.38 & 15.24 \\
 J1412+1635 & 0.454 & 11.1 & 0.39 & 22.5  & 20.91 & 19.3  & 18.88 & 18.57 & 15.5  & 15.21 \\
 J1412+1943 & 0.413 & 10.9 & 0.20   & 21.76 & 20.68 & 19.26 & 19    & 18.75 & 15.79 & 15.57 \\
 J1500+1739 & 0.577 & 10.7 & 0.27 & 19.68 & 19.38 & 19.04 & 18.82 & 18.76 & 15.15 & 14.82 \\
 J1516+1650 & 0.589 & 11.0 & 0.24 & 19.73 & 19.35 & 18.93 & 18.54 & 18.35 & 14.46 & 13.95 \\
 J1049+6433 & 0.454 & 11.4 & 0.38 & 21.79 & 20.36 & 18.78 & 18.35 & 18.13 & 15    & 14.71 \\
 J1528+0126 & 0.403 & 10.9 & 0.22 & 20.3  & 19.78 & 19.03 & 18.62 & 18.23 & 14.53 & 14.06 \\
 J0811+4716 & 0.516 & 11.0 & 0.11 & 21.16 & 20.69 & 19.55 & 19.2  & 18.92 & 15.93 & 15.87 \\
 J0827+2954 & 0.682 & 11.5 & 0.11 & 21.48 & 21.05 & 20.12 & 19.42 & 19.14 & 15.69 & 15.48 \\
 J1613+2834 & 0.449 & 11.0 & 0.24 & 20.26 & 19.76 & 18.94 & 18.69 & 18.42 & 14.84 & 14.25 \\
\hline
\end{longtable} \tablecomments{Average $\log M_{*}/M_{\odot}$ and $\sigma_{\log M_{*}/M_{\odot}}$ were computed in the MCMC fitting routine described above.}

\begin{table}[h!]
    \centering
    \caption{Comparison between our derived stellar masses and those presented in \citet{sell14CSoutflow}.}
    \begin{tabular}{c c c}
    SDSS Name & $\langle \log M_{*}/M_{\odot} \rangle$ & $\log M_{*}/M_{\odot}$ \\
     & (This work) & \citep{sell14CSoutflow} \\
     (1) & (2) & (3) \\
    \hline
    J1506+6131 & $10.3 ^{+0.22} _{-0.15} $ & 10.2 \\
    J0826+4305 & $10.7\pm 0.29$ & 10.8 \\
    J2118+0017 & $10.8 ^{+0.22} _{-0.27}$ & 11.1 \\
    J1558+3957 & $10.6\pm0.24$ & 10.6 \\
    J1613+2834 & $11.0 ^{+0.17} _{-0.24}$ & 11.2 \\
    \end{tabular}
    \label{tab:sell_compare}
    \tablecomments{These derived masses are broadly consistent with one another.}
\end{table}

\begin{figure*}[h!]
\centering
\includegraphics[width=0.8\linewidth]{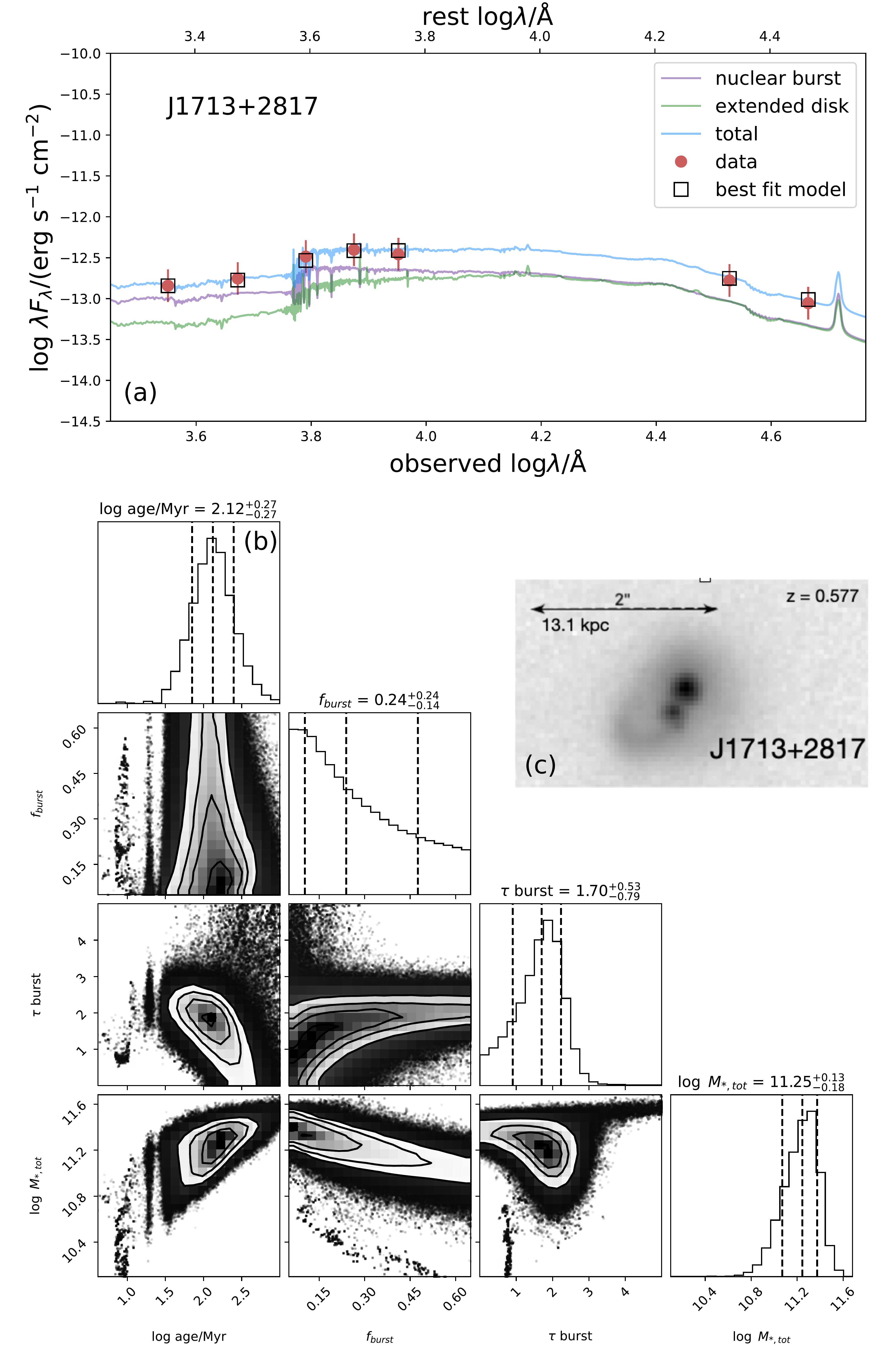}
\caption{\textit{Panel (a)}: Best fit SED for galaxy J01713+2817. The red points and error bars are the observed photometry and $\pm 0.25$ magnitude uncertainty region, respectively. The open black squares are the modeled photometry. The blue, violet, and green curves are the modeled SED for the total galaxy system, nuclear burst, and host galaxy, respectively. \textit{Panel (b)}: Triangle plot of parameter posterior distributions for galaxy J01713+2817. We calculate the mean and covariances of these posterior distributions to model them as 4D-Gaussian distributions. We then randomly draw sets of parameter values from the Gaussian-modeled posterior to construct a mock population of compact starbursts. \textit{Panel (c)}: Galaxy cutout as seen in Figure \ref{fig:cutouts}.}
\label{fig:j1713_app}
\end{figure*}

\begin{figure*}[h!]
\centering
\includegraphics[width=0.8\linewidth]{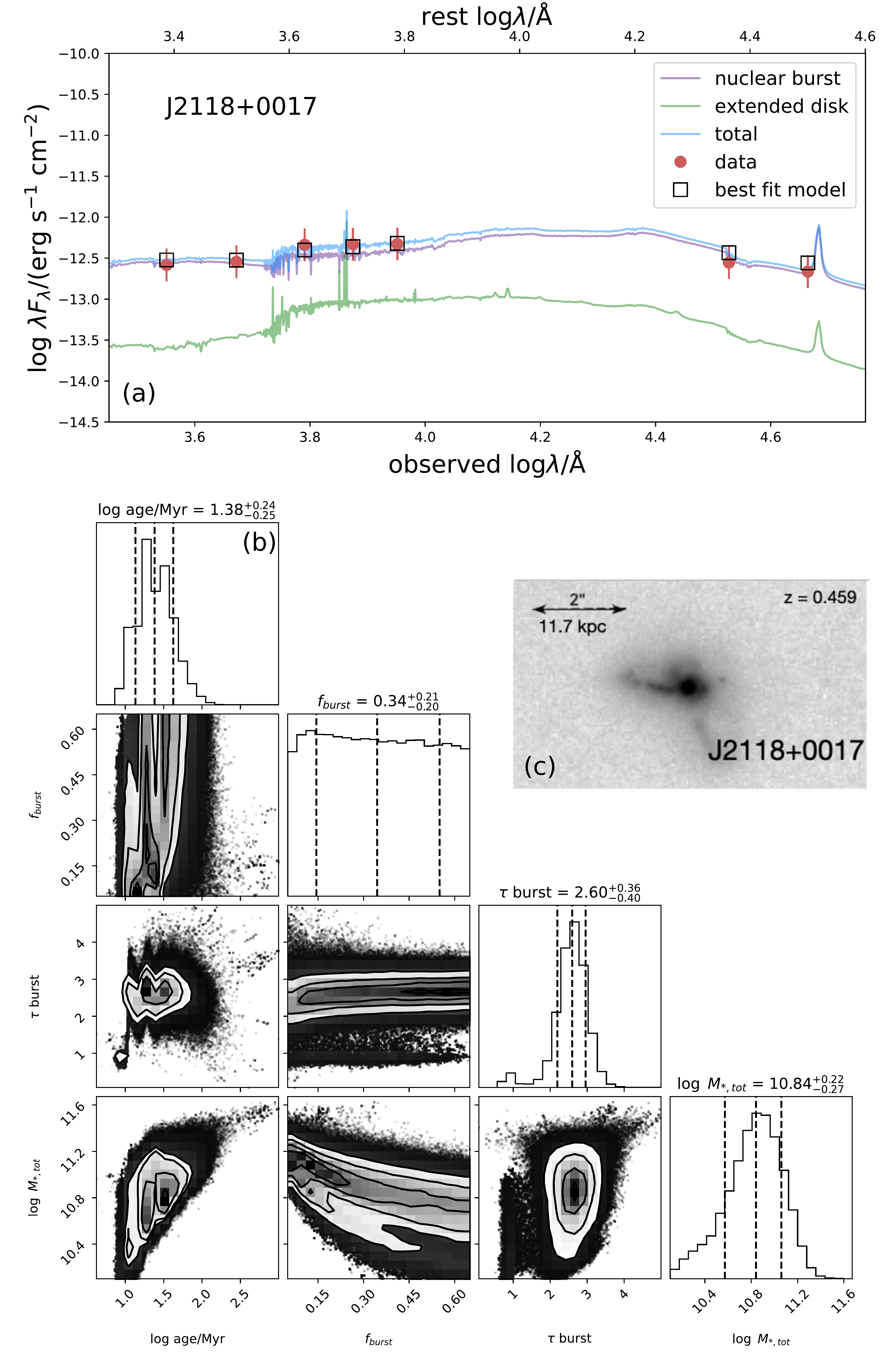}
\caption{\textit{Panel (a)}: Best fit SED for galaxy J2118+0017. The red points and error bars are the observed photometry and $\pm0.25$ magnitude uncertainty region, respectively. The open black squares are the modeled photometry. The blue, violet, and green curves are the modeled SED for the total galaxy system, nuclear burst, and host galaxy, respectively. \textit{Panel (b)}: Triangle plot of parameter posterior distributions for galaxy J2118+0017. We calculate the mean and covariances of these posterior distributions to model them as 4D-Gaussian distributions. We then randomly draw sets of parameter values from the Gaussian-modeled posterior to construct a mock population of compact starbursts. \textit{Panel (c)}: Galaxy cutout as seen in Figure \ref{fig:cutouts}.}
\label{fig:j2118_app}
\end{figure*}

\begin{figure*}[h!]
\centering
\includegraphics[width=0.8\linewidth]{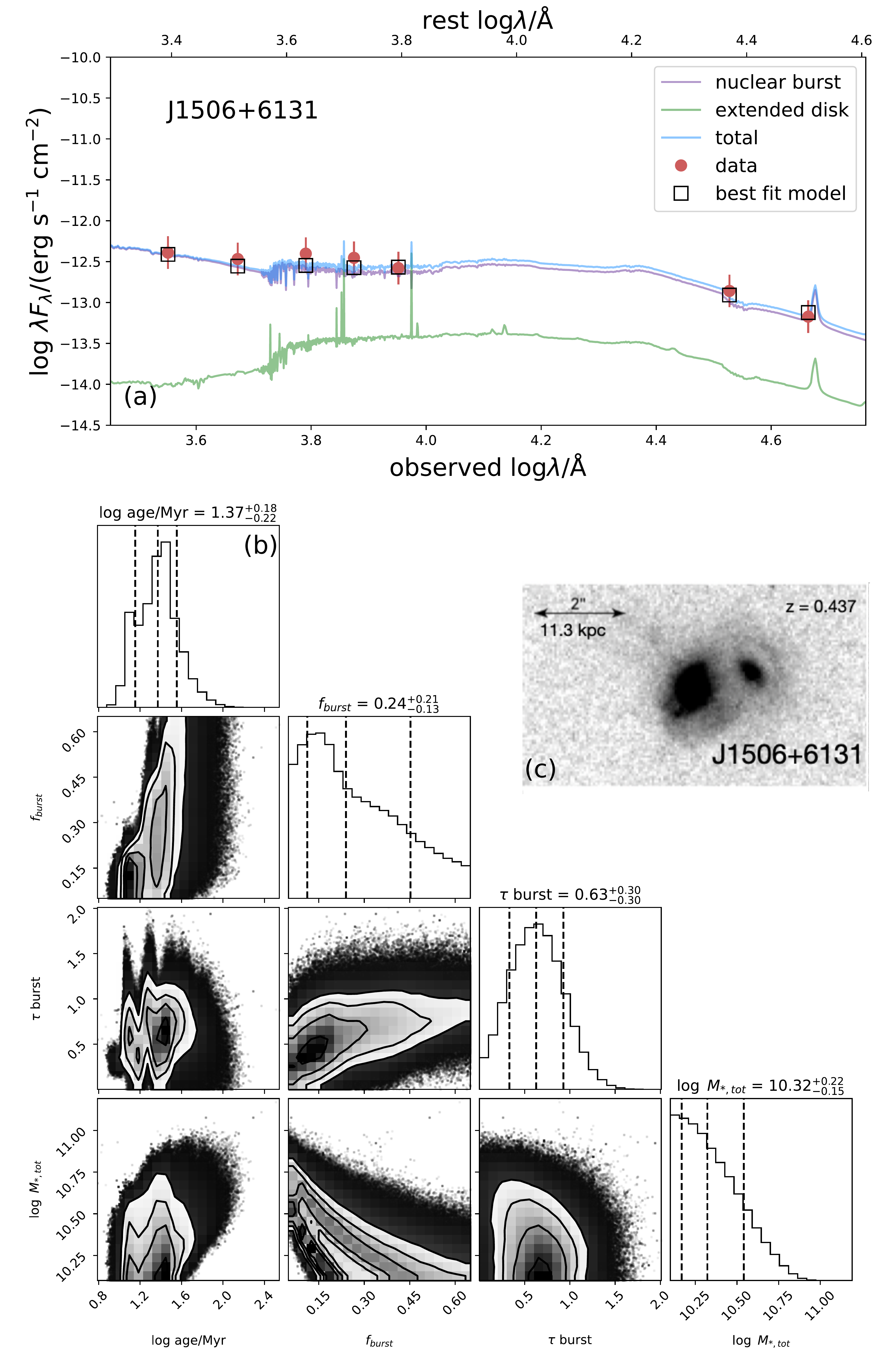}
\caption{\textit{Panel (a)}: Best fit SED for galaxy J1506+6131. The red points and error bars are the observed photometry and $\pm 0.25$ magnitude uncertainty region, respectively. The open black squares are the modeled photometry. The blue, violet, and green curves are the modeled SED for the total galaxy system, nuclear burst, and host galaxy, respectively. \textit{Panel (b)}: Triangle plot of parameter posterior distributions for galaxy J1506+6131. We calculate the mean and covariances of these posterior distributions to model them as 4D-Gaussian distributions. We then randomly draw sets of parameter values from the Gaussian-modeled posterior to construct a mock population of compact starbursts. \textit{Panel (c)}: Galaxy cutout as seen in Figure \ref{fig:cutouts}.}
\label{fig:j1506_app}
\end{figure*}

\begin{figure*}[h!]
\centering
\includegraphics[width=0.8\linewidth]{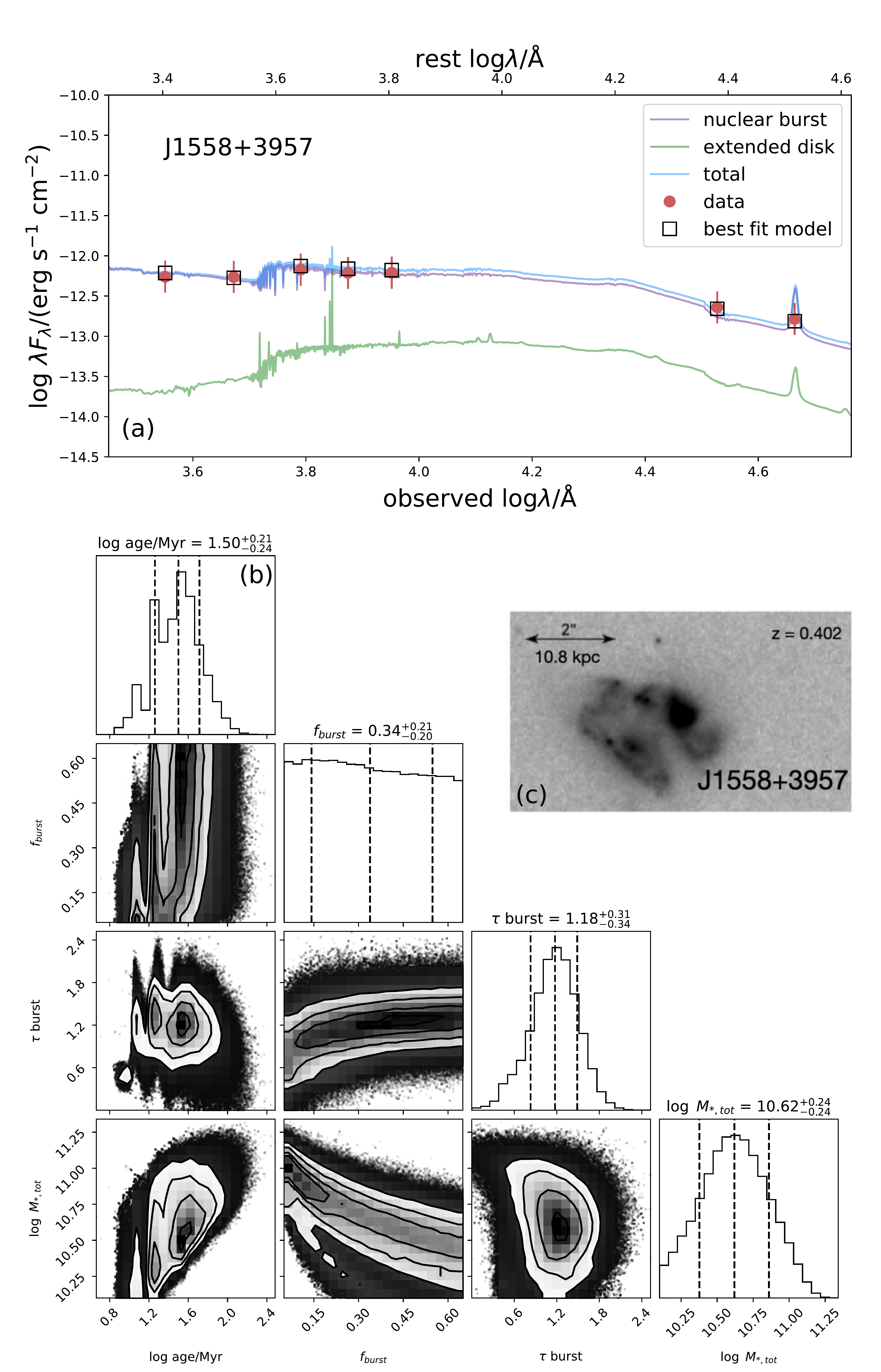}
\caption{\textit{Panel (a)}: Best fit SED for galaxy J1558+3957. The red points and error bars are the observed photometry and $\pm 0.25$ magnitude uncertainty region, respectively. The open black squares are the modeled photometry. The blue, violet, and green curves are the modeled SED for the total galaxy system, nuclear burst, and host galaxy, respectively. \textit{Panel (b)}: Triangle plot of parameter posterior distributions for galaxy J1558+3957. We calculate the mean and covariances of these posterior distributions to model them as 4D-Gaussian distributions. We then randomly draw sets of parameter values from the Gaussian-modeled posterior to construct a mock population of compact starbursts. \textit{Panel (c)}: Galaxy cutout as seen in Figure \ref{fig:cutouts}.}
\label{fig:j1558_app}
\end{figure*}

\begin{figure*}[h!]
\centering
\includegraphics[width=0.8\linewidth]{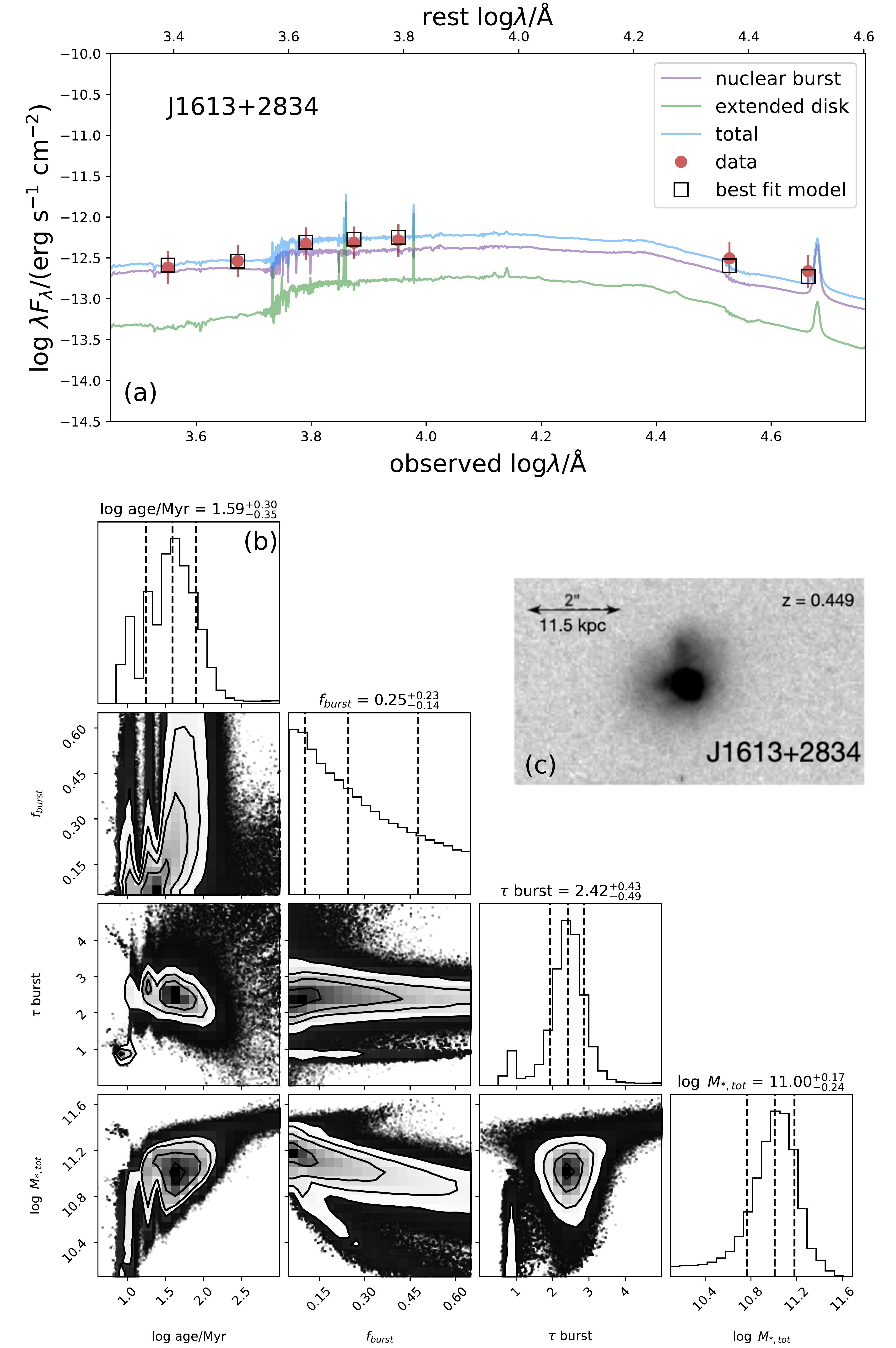}
\caption{\textit{Panel (a)}: Best fit SED for galaxy J1613+2834. The red points and error bars are the observed photometry and $\pm$0.25 magnitude uncertainty region, respectively. The open black squares are the modeled photometry. The blue, violet, and green curves are the modeled SED for the total galaxy system, nuclear burst, and host galaxy, respectively. \textit{Panel (b)}: Triangle plot of parameter posterior distributions for galaxy J1613+2834. We calculate the mean and covariances of these posterior distributions to model them as 4D-Gaussian distributions. We then randomly draw sets of parameter values from the Gaussian-modeled posterior to construct a mock population of compact starbursts. \textit{Panel (c)}: Galaxy cutout as seen in Figure \ref{fig:cutouts}.}
\label{fig:j1613_app}
\end{figure*}
\end{document}